\documentclass[prx,twocolumn,superscriptaddress,secnumarabic,balancelastpage,amsmath,amssymb]{revtex4-1}
\usepackage{amssymb}
\usepackage{amsmath}
\usepackage[dvipsnames]{xcolor}
\usepackage{chngpage}
\usepackage{lgrind}        
\usepackage{color}         
\usepackage{graphics}      
\usepackage[pdftex]{graphicx}      
\usepackage{longtable}     
\usepackage{epsf}          
\usepackage{bm}            
\usepackage{asymptote}     
\usepackage{thumbpdf}
\usepackage{verbatim}
\usepackage{url}
\usepackage{MnSymbol}
\usepackage{physics}
\usepackage[colorlinks=true]{hyperref} 
\usepackage{enumitem}	

\usepackage{float}
\usepackage{times}
\usepackage{color}
\usepackage{verbatim}
\usepackage{soul,xcolor}
\setstcolor{red}
\usepackage{colortbl}

\usepackage{tabularx}

\makeatletter
\AtBeginDocument{
\g@addto@macro{\appendix}{\renewcommand{\p@subsection}{\@Alph\c@section}}
}
\makeatother

\begin{document}

\title{Robust Dynamic Hamiltonian Engineering of Many-Body Spin Systems}

\affiliation{Department of Physics, Harvard University, Cambridge, Massachusetts 02138, USA}
\affiliation{School of Engineering and Applied Sciences, Harvard University, Cambridge, Massachusetts 02138, USA}
\affiliation{Department of Physics, University of California Berkeley, Berkeley, California 94720, USA}

\author{Joonhee Choi$^{1,2}$}
\thanks{These authors contributed equally to this work} 
\author{Hengyun Zhou$^{1}$}
\thanks{These authors contributed equally to this work} 
\author{Helena S. Knowles$^{1}$}
\author{Renate Landig$^{1}$}
\author{Soonwon Choi$^{3}$} 
\author{Mikhail D. Lukin$^{1}$}
\email{lukin@physics.harvard.edu}


\begin{abstract}
We introduce a new approach for the robust control of quantum dynamics of strongly interacting many-body systems. Our approach involves the design of periodic global control pulse sequences to engineer desired target Hamiltonians that are robust against disorder, unwanted interactions and pulse imperfections. It utilizes a matrix representation of the Hamiltonian engineering protocol based on time-domain transformations of the Pauli spin operator along the quantization axis.
This representation allows us to derive a concise set of algebraic conditions on the sequence matrix to engineer robust target Hamiltonians, enabling the simple yet systematic design of pulse sequences.
We show that this approach provides an efficient framework to (i) treat \textit{any} secular many-body Hamiltonian and engineer it into a desired form, (ii) target dominant disorder and interaction characteristics of a given system, (iii) achieve robustness against imperfections, (iv) provide optimal sequence length within given constraints, and (v) substantially accelerate numerical searches of pulse sequences.
Using this systematic approach, we develop novel sets of pulse sequences for the protection of quantum coherence, optimal quantum sensing and quantum simulation.
Finally, we experimentally demonstrate the robust operation of these sequences in a system with the most general interaction form.
\end{abstract}
\maketitle

\section{Introduction and Motivation}
The ability to control and manipulate the dynamics of a quantum system in a robust fashion is key to many quantum technologies. In particular, the use of periodic control pulses, also known as Floquet driving, has emerged as a ubiquitous tool for the control and engineering of quantum dynamics~\cite{haeberlen1968coherent,vandersypen2005nmr,eckardt2017colloquium,oka2018floquet,bukov2015universal,goldman2014periodically,poudel2015dynamical},
with applications in protecting quantum coherence from environmental noise~\cite{hahn1950spin,carr1954effects,meiboom1958modified,gullion1990new,viola1999dynamical,khodjasteh2005fault,viola2005random,uhrig2007keeping,khodjasteh2009dynamical,biercuk2009optimized,du2009preserving,alvarez2010performance,west2010near,de2010universal,ryan2010robust,khodjasteh2013designing,suter2016,burum1979analysis,cory1990time,iwamiya1993application,naydenov2011dynamical,genov2017arbitrarily}
and frequency-selective quantum sensing~\cite{schirhagl2014nitrogen,degen2017quantum,taylor2008high,degen2008scanning,bylander2011noise,hall2010ultrasensitive,naydenov2011dynamical,alvarez2011measuring,de2011single,pham2012enhanced,norris2016qubit,frey2017application,rose2018high,fiderer2018quantum,lang2015dynamical}.
Periodic control can also be employed to engineer the interactions between qubits, even when only global control is available, enabling the study of out-of-equilibrium phenomena, such as dynamical phase transitions and quantum chaos, and the observation of novel phases of matter such as discrete time crystals~\cite{lindner2011floquet,jiang2011majorana,heyl2018dynamical,dalessio2016from,garttner2017measuring,khemani2016phase,else2016floquet,von2016absolute,yao2017discrete,choi2017observation,zhang2017observation,sacha2018time,nandkishore2015many,abanin2018many,alvarez2015localization,wei2018exploring,wei2018emergent,ho2017critical,choi2019probing}.

The key tool to engineer the dynamics of periodically driven systems is average Hamiltonian theory (AHT). This technique has been particularly successful in nuclear magnetic resonance (NMR), where periodic driving protocols enable the suppression of unwanted evolution due to both disorder and interactions, effectively preserving quantum coherence and enabling high-resolution NMR spectroscopy and magnetic resonance imaging (MRI)~\cite{slichter2013principles,mehring2012principles,levitt2001,sorensen1984product,rhim1971time,drobny1978fourier,burum1979analysis,shaka1983improved,baum1985multiple,shaka1988iterative,tycko1990zero,cory1990time,lee1995efficient,hohwy1999fivefold,carravetta2000symmetry,takegoshi1985magic,rose2018high,lee1965nuclear,vinogradov1999high,iwamiya1993application,sakellariou2000homonuclear}.

However, conventional control pulse sequences are generally optimized for solid-state nuclear spin systems where dipolar interactions dominate. In particular, these sequences are often not applicable to other quantum systems, such as electronic spin ensembles or arrays of coupled qubits, where either on-site disorder dominates or interactions have a more general form~\cite{kucsko2018critical, mohammady2018low}.

Furthermore, periodic driving schemes are often vulnerable to perturbations caused by inhomogeneities of individual spins in the system, non-ideal finite pulse duration effects, as well as imperfect spin state manipulation. While there exist many pulse sequences that retain robustness to some of these control imperfections~\cite{rhim1974analysis,burum1979analysis,cory1990time}, a systematic framework to treat these errors in a general setting of interest is still lacking, hindering the customized design of pulse sequences optimally adapted for various applications across different experimental platforms.

In this work, we introduce a novel framework to systematically address these challenges and efficiently design robust, self-correcting pulse sequences for dynamic Hamiltonian engineering in interacting spin ensembles using only global control~\cite{hayes2014programmable,ajoy2013quantum,choi2017dynamical,okeeffe2019hamiltonian,haas2019engineering,attar2019hamiltonian}. 
Such globally controlled spin ensembles are naturally realized in various systems~\cite{waugh1968approach,wei2018exploring,kucsko2018critical,tyryshkin2003electron,blatt2012quantum,jurcevic2014quasiparticle,bohnet2016quantum,zhang2017observation,labuhn2016tunable,bernien2017probing}. We demonstrate both theoretically and experimentally that our approach has immediate applications ranging from dynamical decoupling and quantum metrology to quantum simulation.

\begin{figure}
\begin{center}
\includegraphics[width=\columnwidth]{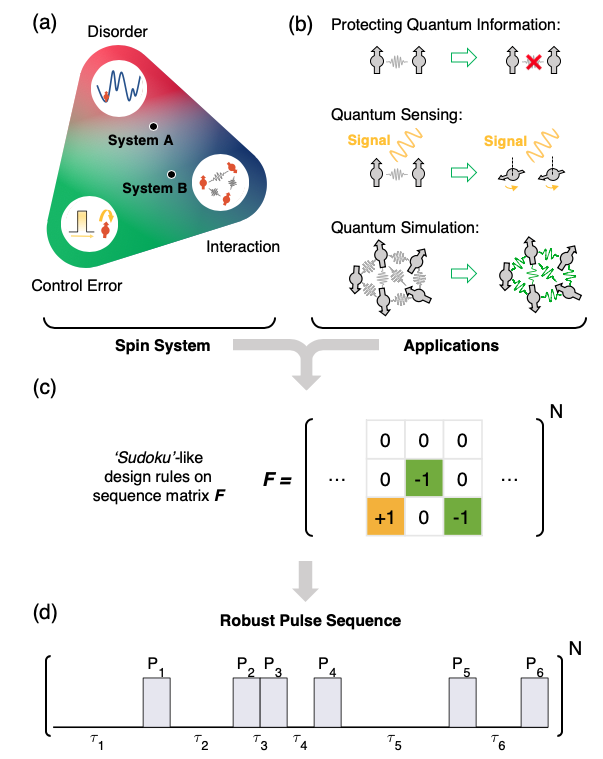}
\caption{{\bf Optimal pulse sequence design for robust dynamic Hamiltonian engineering.} (a) Illustration of the interplay between disorder, interactions and control errors in different quantum systems, with examples of a disorder-dominated system (System A) and an interaction-dominated system (System B). (b) Applications of driven quantum many-body systems. (c) Our Hamiltonian engineering approach is based on `Sudoku'-puzzle-like design rules, imposed on the matrix $F$ that represents a periodic pulse sequence. (d) Resulting robust periodic pulse sequence, optimized for a target application with system-targeted design. The sequence, characterized by $n$ finite-duration pulses $\{P_1,\cdots,P_n\}$ with free evolution intervals between the pulses $\{\tau_1,\cdots,\tau_n\}$, is periodically applied to a system to dynamically engineer the Hamiltonian.} 
\label{fig:fig1}
\end{center}
\end{figure}

Our approach is based on a simple matrix representation of pulse sequences that allows for their analysis and design in a straightforward fashion, using intuitive algebraic conditions. This matrix describes the interaction-picture transformations of the $S^z$ operator, the Pauli spin operator along the quantization axis, which can also be visualized in a very intuitive way.
Crucially, we show that by decomposing all pulses into $\pi/2$-pulse building blocks, this representation not only gives the effective leading-order average Hamiltonian describing the driven spin dynamics, but also provides a concise description of dominant imperfections arising from non-ideal, finite-duration pulses and rotation angle errors.
More specifically, we show that (i) the suppression or tuning of on-site potential disorder, (ii) the decoupling or engineering of spin-spin interactions, and (iii) the robustness of the pulse sequence against control imperfections, can all be extracted directly from our representation and algebraic conditions.
The algebraic conditions also analytically provide the minimum number of pulses required to realize a target application, thus ensuring minimal sequence length under given constraints.
This approach thus allows the incorporation of Hamiltonian engineering requirements in the presence of imperfections, enabling the versatile construction of sequences designed for a particular quantum application and tailored to the detailed properties of the experimental system at hand (see Fig.~\ref{fig:fig1}).
 
Specifically, we use our formalism to \textit{protect quantum information} and benchmark the performance of two sequences with different design considerations, each suited for systems in different regimes of competing disorder and interaction energy scales.
We also utilize our framework to design pulse sequences for robust and optimal \textit{quantum sensing}, where our method provides a generalized picture of AC field sensing protocols in which an external AC field in the lab frame translates into an effective vector DC field in the driven spin frame. Combining optimal choices of the effective DC sensing field and initialization/readout directions with coherence time extensions through disorder and interaction suppression, this approach can lead to high sensitivity magnetometry beyond the limit imposed by spin-spin interactions, as we show in a separate manuscript~\cite{zhou2019quantum}.
We then further apply our framework to \textit{quantum simulation} and engineer the bare system Hamiltonian to a desired target form, providing a new avenue to study many-body dynamics over a wide range of tunable parameters with different types of interactions and disorder.
Finally, we experimentally demonstrate our results in a disordered, dipolar-interacting nitrogen-vacancy (NV) center ensemble in diamond with the most general form of interactions.

The main advances enabled by our approach include:
\begin{enumerate}
\item \textbf{Robustness}: We show that all types of average Hamiltonian effects, including errors resulting from pulse imperfections [Sec.~\ref{sec:imperfect}], can be readily incorporated as concise algebraic conditions on the transformation properties of the $S^z$ Pauli spin operator in the interaction picture. This leads to a simple recipe for sequence robustness by design.

\item \textbf{Generality}: Our approach is applicable to generic two-level spin ensembles in a strong quantizing field, as typically found in most experimental quantum many-body platforms such as solid-state electronic and nuclear spin ensembles, trapped ions, molecules, neutral atoms, or superconducting qubits. Our framework covers on-site disorder, various two-body interaction types such as Ising interactions and spin-exchange interactions, as well as complex \textit{three-body} interactions [Sec.~\ref{sec:multibody}].

\item \textbf{Flexibility}: The flexibility of our approach allows Hamiltonian engineering that takes the energy hierarchy into account, which can be tailored to specific physical systems exhibiting different relative strengths between disorder, interactions, and control errors [Sec.~\ref{sec:decouple}]. This enables the development of pulse sequences designed for disorder-dominated systems, beyond the typical NMR setting.

\item \textbf{Efficiency}:  Using simple algebraic conditions, we can find the shortest possible sequence length required to achieve a target Hamiltonian [Sec.~\ref{sec:Shortest}]. In addition, we use combinatorial analysis to demonstrate the necessity of composite pulse structures for efficient sensing, and provide optimized sequences that achieve maximum sensitivity to external signals [Sec.~\ref{sec:senseopt}]. The algebraic conditions also substantially improve numerical searches of pulse sequences by constraining the search space to a set of good pulse sequences [Sec.~\ref{sec:numerics}].
\end{enumerate}

The paper is organized as follows: In Secs.~\ref{sec:perfect} and~\ref{sec:imperfect}, we provide the theoretical framework for systematic pulse design. This is extended to higher-order and more complex, multi-body interacting Hamiltonians via analytical and numerical approaches in Sec.~\ref{sec:HighOrder}. In Secs.~\ref{sec:decouple},~\ref{sec:sensing} and~\ref{sec:simulation}, we present system-targeted sequence design for the applications of dynamical decoupling, quantum sensing and quantum simulation, respectively.
Finally, Sec.~\ref{sec:experiment} presents the experimental demonstration of our results to dynamical decoupling, with a particular focus on the broad applicability under different forms of the Hamiltonian. We conclude with a further discussion and outlook of the framework in Sec.~\ref{sec:conclusion}.

\section{General Formalism and Frame Representation}
\label{sec:perfect}
We start by introducing a simple representation of pulse sequences based on the rotations of the spin frame in the interaction picture: Instead of illustrating a sequence by the applied spin-rotation pulses, we describe it by specifying how the $S^z$ spin operator is rotated by the applied pulses in the interaction picture (also known as the toggling-frame picture). This method provides a one-to-one correspondence with the average Hamiltonian of the system and is an extension of the method presented in Ref.~\cite{mansfield1971symmetrized}, also closely related to control matrices~\cite{green2013arbitrary,paz-silva2014general} and vector modulation functions~\cite{lang2017enhanced,lang2019nonvanishing,schwartz2018robust,wang2019randomization}. Note however that the form of the Hamiltonian is limited in these existing papers, and robust decoupling rules in the interacting regime have not been derived. We efficiently depict this $S^z$ operator evolution using a simple matrix, and show that this direct link to the average Hamiltonian is valid for any system under a strong quantizing field. In addition to its simplicity in describing the decoupling performance for the case of ideal, instantaneous pulses, this representation also allows the formulation of concise criteria to treat pulse imperfections, as will be discussed in Sec.~\ref{sec:imperfect}.

\begin{figure*}
\begin{center}
\includegraphics[width=2\columnwidth]{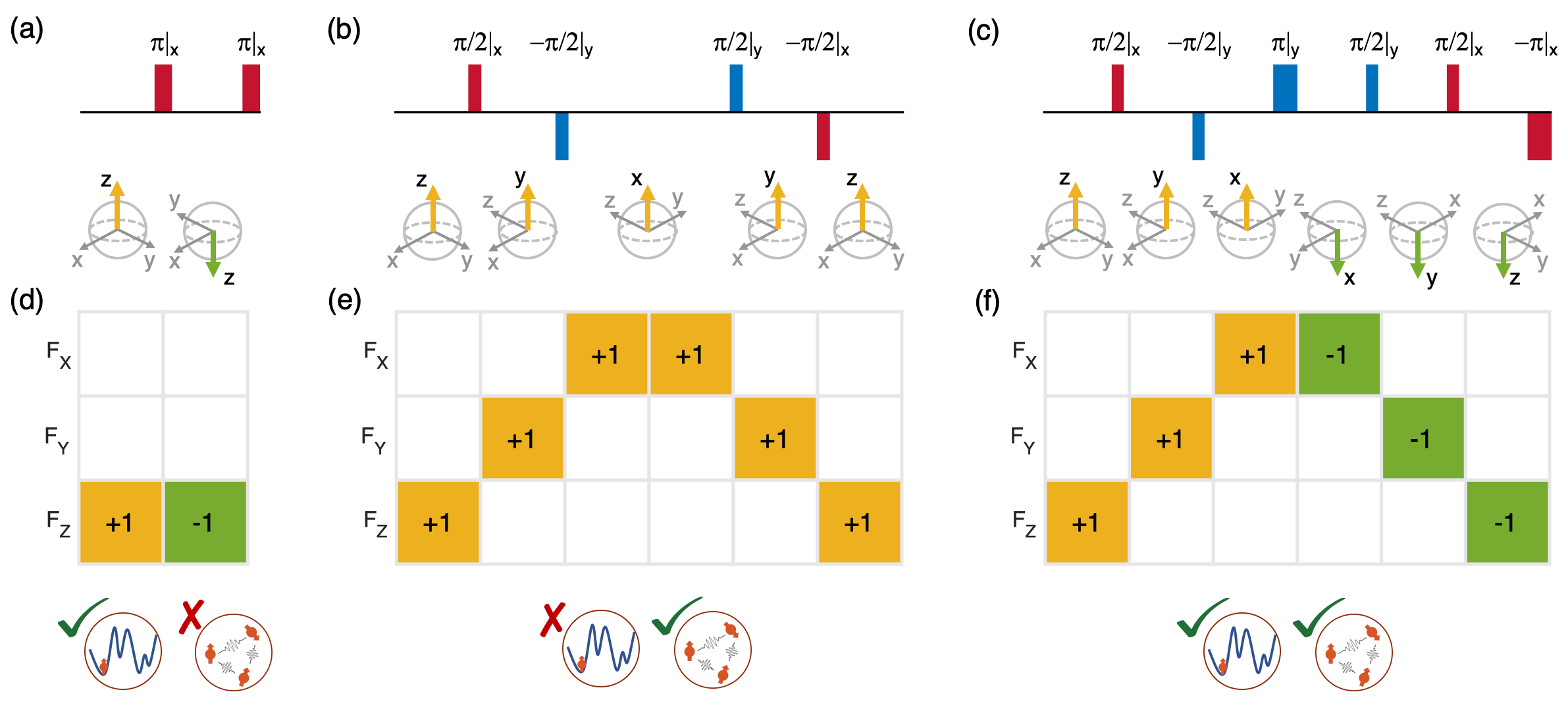}
\caption{{\bf Efficient representation of pulse sequences}. (a)-(c) Conventional illustrations of pulse sequences performing $\pm\pi$ and $\pm\pi/2$ rotations around the $\hat{x}$-axis (red) and $\hat{y}$-axis (blue). The spheres below the sequences describe time-domain transformations of spin frames (spin operators instead of states) in the interaction picture, periodically rotated by pulses from each sequence. In our framework, we only focus on the rotation of the $S^z$ operator in the time domain, whose orientation is highlighted as yellow and green arrows for positive and negative axis directions, respectively. The rotations employ the standard convention of a right-handed coordinate system. (a) CPMG sequence designed to decouple spins from on-site disorder. (b) WAHUHA sequence designed to suppress spin-spin interactions and (c) Echo+WAHUHA sequence designed to cancel both disorder and interaction effects. (d)-(f) Efficient matrix-based representation of each periodic pulse sequence. The 3-by-$n$ matrix ${\bf F} = [F_{\mu,k}]$ is employed to describe such time-domain $S^z$ operator transformations in a simple form; For example, $F_{\mu,k}$, a nonzero matrix element at $(\mu,k)$, means that $S^z$ transforms to $S^\mu$ in the $k$-th frame. The bottom insets illustrate different decoupling characteristics of each sequence, where checks and crosses indicate success and failure in suppressing disorder (left circle) and interaction (right circle) effects, respectively.}
\label{fig:fig2}
\end{center}
\end{figure*}

\subsection{Frame Representation}
\label{sec:framerep}
The dynamics of periodically driven systems can be described and analyzed using AHT~\cite{haeberlen1968coherent} (see Appx.~\ref{sec:AHT} for a detailed review of AHT). In particular, for a pulse sequence consisting of $n$ spin-rotation pulses \{$P_{1,\cdots,n}$\}, the leading-order average Hamiltonian, $H_\text{avg}$, is a simple weighted average of the toggling-frame Hamiltonians
\begin{align}
H_\text{avg} =\frac{1}{T} \sum_{k=1}^n \tau_k \tilde{H}_k,
\label{eq:Haverage}
\end{align}
where $\tau_k$ is the pulse spacing between the $k-1$-th and $k$-th control pulses $P_{k-1}$ and $P_k$, $\tilde{H}_k=(P_{k-1}\cdots P_1)^\dagger H_s (P_{k-1}\cdots P_1)$ is the toggling-frame Hamiltonian that governs spin dynamics during the $k$-th evolution period, $\tau_k$, in the interaction picture, and $H_s$ is the internal system Hamiltonian.

Here, we present a convenient alternative method to calculate the leading-order average Hamiltonian, utilizing our toggling-frame sequence representation~\cite{mansfield1971symmetrized}. Our representation is based on the time-domain transformations of a single-body $S^z$ spin operator in the interaction picture. As shown in Sec.~S1A~\cite{SM}, the representation works for general two-level system Hamiltonians under the rotating wave approximation in a strong quantizing field (secular approximation). Physically, this corresponds to the common situation, realized in almost all experimental platforms, in which energetic considerations require all interaction terms to conserve the total magnetization along the quantization axis $\hat{z}$, which can also be written as $[H_s, S^z_\text{tot}] = 0$, where $S^z_\text{tot} = \sum_i S_i^z$ is the total spin projection operator along the $\hat{z}$-axis. Thus, our framework is widely applicable to different experimental systems, including both ordered and disordered systems, and systems with different types of interactions, including Ising~\cite{bernien2017probing,zhang2017observation}, spin-exchange~\cite{kucsko2018critical,mohammady2018low}, dipolar~\cite{waugh1968approach}, and even exotic three-body interactions~\cite{buchler2007three,mezzacapo2014many,chancellor2017circuit}.

To introduce our framework in detail, let us first consider two-body interaction Hamiltonians with on-site disorder. The most general form of such a Hamiltonian is (Sec.~S1A~\cite{SM})
\begin{align}
H_s&=H_\text{dis}+H_\text{int} \nonumber\\&=\sum_i h_i S_i^z+ \sum_{ij}\left[J^{I}_{ij} S_i^zS_j^z + J^S_{ij} \qty(S_i^xS_j^x+S_i^yS_j^y)\nonumber\right.\\&\hspace{2.2cm} \left.+J^A_{ij}\qty(S_i^xS_j^y-S_i^yS_j^x)\right],
\label{eq:genHam}
\end{align}
where the first term $H_\text{dis}$ is the on-site disorder Hamiltonian and the second term $H_\text{int}$ is a generic two-body interaction Hamiltonian, $h_i$ is a random on-site disorder strength, $\{S^x_i,S^y_i,S^z_i\}$ are spin-1/2 operators, and $J^I_{ij}, J^S_{ij}, J^A_{ij}$ are arbitrary interaction strengths for the Ising interaction and the symmetric and anti-symmetric spin-exchange interactions, respectively.
According to AHT, a pulse sequence periodically applied to the system can engineer this into a new Hamiltonian, dictated by the control field that dynamically manipulates the spins (see Appx.~\ref{sec:AHT}).

In our framework, the control field is assumed to be a time-periodic sequence of short pulses, with each pulse constructed out of $\pi/2$-rotation building blocks around the $\hat{x}$, $\hat{y}$ axes [Fig.~\ref{fig:fig1}(d)].
In this setting, let us consider the interaction-picture transformations of the $S^z$ operator: $\tilde{S}^z(t)=U_c^\dagger(t) S^zU_c(t)$, where $U_c(t)$ is the global unitary spin rotation due to the control field. We will assume in this section that the pulses are perfect and infinitely short. In such a case, the $+S^z$ operator transforms into $\pm S^{x,y,z}$ operators, depending on the rotation angles and axes of the pulses. Hence, the effect of the pulse sequence is a rotation of $S^z$ in time in a discrete fashion, and the transformation trajectory in the toggling frame can be identified as
\begin{align}
	\tilde{S}^z(t) &= (P_{k-1} \cdots P_1)^\dagger S^z (P_{k-1} \cdots P_1) \nonumber \\
	&=\sum_\mu F_{\mu,k} S^\mu, \quad \text{for $t_{k-1} < t < t_k$}.
\label{eq:Sz}
\end{align}
Here, $P_k$ is the \textit{global} spin rotation performed right after the $k$-th toggling frame, $t_k = \sum_{j=1}^k \tau_j$ with $t_0$ = 0, and ${\bf F} =  [F_{\mu,k}] = [\vec{F}_x; \vec{F}_y; \vec{F}_z]$ is a $3 \times n$ matrix containing elements 0 and $\pm1$. The matrix elements $F_{\mu,k}$ can be explicitly calculated as
\begin{align}
    F_{\mu,k} = 2\Tr[S^\mu \tilde{S}^z_k],
\end{align}
with $\tilde{S}^z_k = \tilde{S}^z(t)$ for $t_{k-1} < t < t_k$. Intuitively, a nonzero element $F_{\mu,k}$ indicates that the initial $S^z$ operator transforms into $S^\mu$ for the duration of the free evolution interval $\tau_k$, with its sign determined by $F_{\mu,k}$. Additionally, the time duration of each toggling frame is specified by the frame-duration vector $\boldsymbol{\tau} = [\tau_1, \tau_2, \cdots, \tau_n]$.

This representation is illustrated for three pulse sequences in Fig.~\ref{fig:fig2}. The CPMG sequence~\cite{carr1954effects,meiboom1958modified,gullion1990new}, consisting of equidistant $\pi$ pulses to suppress on-site disorder, can be represented as 
\begin{align}
\begin{pmatrix}
{\bf F}\\
\boldsymbol{\tau}
\end{pmatrix}_\text{CPMG}
=\begin{pmatrix}
0 & 0 \\
0 & 0 \\
+1 & -1 \\
\tau& \tau\\
\end{pmatrix},\nonumber
\end{align}
since the first $+S_z$ is flipped to $-S_z$ by a $\pi$ pulse [Fig.~\ref{fig:fig2}(a,d)]. Similarly, the WAHUHA sequence, consisting of four $\pi/2$ pulses~\cite{waugh1968approach}, is shown in Fig.~\ref{fig:fig2}(b,e). The matrix clearly shows how the spin operator rotates over time, cycling through all three axes to cancel dipole-dipole interactions~\footnote[1]{
In our algebraic conditions, we use the convention where each free evolution time is immediately followed by a pulsed rotation. For base pulse sequences that end with a free evolution in the original $\hat{z}$-axis without any following pulse, as in the WAHUHA sequence case [Fig.~\ref{fig:fig2}(b,e)], we move the final free-evolution block to the beginning of the pulse sequence representation and combine it with the first frame, before applying our algebraic conditions for robustness [Tab.~\ref{tab:summary}].}. Finally, we present a sequence that combines the ideas of WAHUHA and CPMG to echo out disorder while symmetrizing interactions, as depicted in Fig.~\ref{fig:fig2}(c,f).

The representation thus far uniquely specifies the toggling-frame $\tilde{S}^z$ orientation after each instantaneous pulse. However, the rotation axis of $\pi$ pulses is not yet uniquely specified. To address this, we decompose all pulses into $\pi/2$ building blocks, and specify intermediate frames for pulses of rotation angles larger than $\pi/2$. Such a $\pi/2$-pulse decomposition also simplifies the analysis of finite pulse duration effects, which we discuss in Sec. \ref{sec:imperfect}. As shown in Fig.~\ref{fig:fig3}(a), a $\pi$ pulse is then split into two $\pi/2$ pulses with zero time separation, where the first $\pi/2$ pulse along the $\hat{x}$-axis rotates $+S^z$ into the intermediate frame $+S^y$ and the second $\pi/2$ pulse along the same axis brings $+S^y$ into $-S^z$.
For example, the CPMG sequence involving $\pi$ pulses can be represented as
 \begin{align}
 \begin{pmatrix}
{\bf F}\\
\boldsymbol{\tau}
\end{pmatrix}_\text{CPMG}
=\begin{pmatrix}
0 & 0 & 0 & 0\\
0 & +1 & 0 & -1\\
+1 & 0 & -1 & 0\\
\tau & 0 & \tau & 0
\end{pmatrix},
\nonumber
\end{align}
which now unambiguously specifies the rotation axes for the $\pi$ pulses. Note that the sequence contains zeros in its frame-duration vector, serving to indicate the intermediate frames, and in the following we shall indicate them with narrow lines in the pictorial representation (see Fig.~\ref{fig:fig3}). The use of such intermediate frames also allows a natural description of composite $\pi/2$ pulse structures, which will play an important role in robust quantum sensing sequences [Sec.~\ref{sec:sensing}].

One key advantage of our pulse sequence representation $\{{\bf F}, \boldsymbol{\tau} \}$ is that we can now conveniently obtain the engineered $H_\text{avg}$ [Eq.~(\ref{eq:Haverage})] from any many-body Hamiltonian of the form Eq.~(\ref{eq:genHam}). 
More specifically, we find that the weighted row-sums and weighted absolute row-sums (which are equivalent to row square sums, since each element $F_{\mu,k}$ takes on values $\{0,\pm1\}$) of the sequence matrix,
\begin{align}
K_\mu&=\frac{1}{T}\sum_{k=1}^n F_{\mu,k} \tau_k,\label{eq:rowsum}\\
L_\mu&=\frac{1}{T}\sum_{k=1}^n \qty|F_{\mu,k}| \tau_k,
\label{eq:absrowsum}
\end{align}
fully specify the generalized formula for the average Hamiltonian $H_\text{avg}$ as a result of the following toggling-frame transformations of two-body Ising, symmetric exchange and anti-symmetric exchange interaction Hamiltonians:
\begin{align}
S_i^z S_j^z  &\rightarrow \sum_{\mu} |F_{\mu,k}| S^\mu_i S^\mu_j  \\
S_i^x S_j^x + S_i^y S_j^y  &\rightarrow \sum_{\mu} (1-|F_{\mu,k}|) S^\mu_i S^\mu_j  \\
S_i^x S_j^y - S_i^y S_j^x  &\rightarrow \sum_{\mu} F_{\mu,k} (\vec{S}_i \times \vec{S}_j)^\mu.
\end{align}
These expressions can be intuitively understood by examining how the $S^z$ operator is transformed, and using an analogy between the anti-symmetric interaction form and cross products (see Appx.~\ref{suppsec:aveideal} for detailed derivations).
Using $K_\mu, L_\mu$ defined above, we can thus write the leading-order average Hamiltonian, $H_\text{avg} = H_\text{avg}^\text{dis} + H_\text{avg}^\text{int}$, as
\begin{align}
H_\text{avg}^\text{dis}&=\sum_{i,\mu} h_i S_i^\mu \cdot K_\mu, \label{eq:Havedis} \\
H_\text{avg}^\text{int}&=\sum_{ij,\mu} J^I_{ij} S_i^\mu S_j^\mu\cdot L_\mu \label{eq:HaveIsing} \\
				&+\sum_{ij,\mu} J^S_{ij} S_i^\mu S_j^\mu\cdot \left( 1-L_\mu \right)  \label{eq:HaveSym}  \\
				&+\sum_{ij,\mu} J^A_{ij} (\vec{S}_i \times \vec{S}_j)^\mu\cdot K_\mu.  \label{eq:HaveAntiSym}
\end{align}

\renewcommand{\figurename}{Table}
\setcounter{figure}{0}
\begin{figure*}
\begin{center}
\includegraphics[width=2\columnwidth]{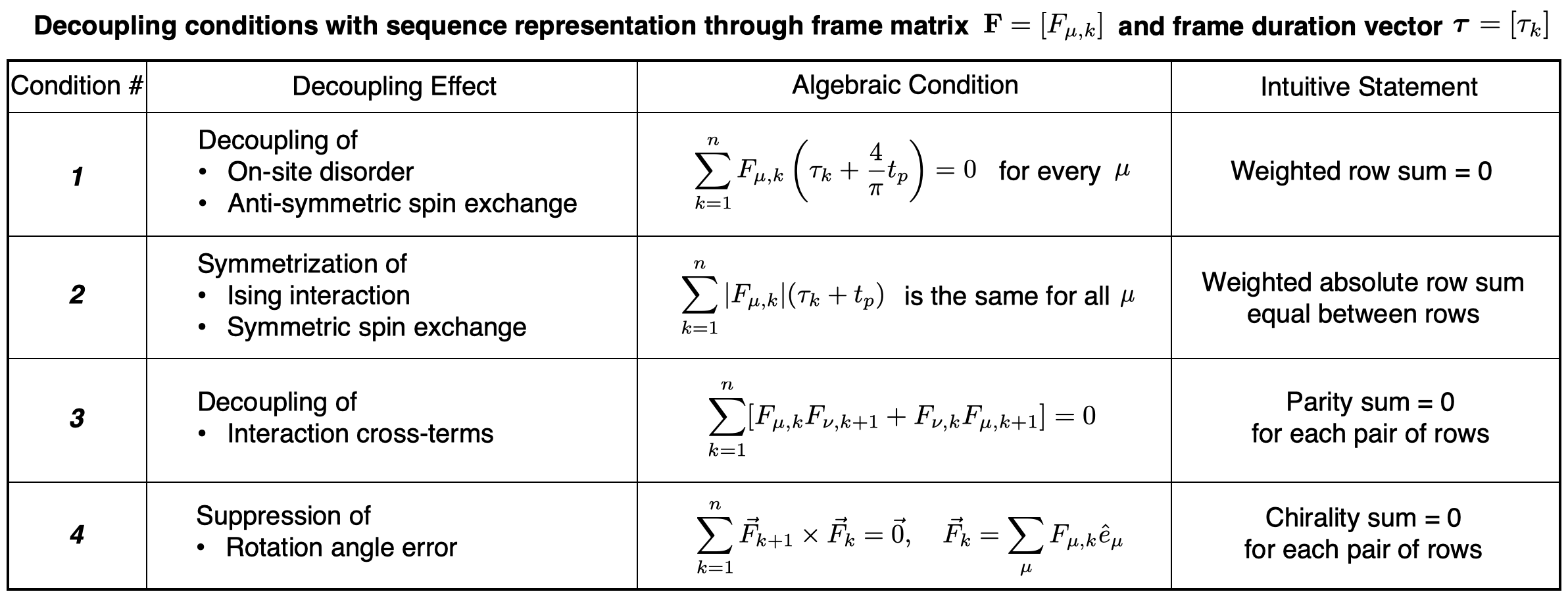}
\caption{Summary of robust dynamical decoupling conditions. A periodic pulse sequence consisting of $n$ $\pi/2$ pulse building blocks is represented by a 1-by-$n$ frame duration vector $\boldsymbol{\tau} = [\tau_k]$ and a 3-by-$n$ frame matrix ${\bf F} = [F_{\mu,k}] = [\vec{F}_x; \vec{F}_y; \vec{F}_z]$, corresponding to the toggling-frame $S^z$ operator in different evolution blocks [Sec.~\ref{sec:framerep}]. Based on this representation, the conditions for dynamical decoupling and fault-tolerance against leading-order imperfections can be phrased as intuitive statements on ${\bf F}$, with precise algebraic conditions as listed. Above, $t_p$ is the duration of a $\pi/2$ pulse, $\hat{e}_\mu$ is the unit vector along axis $\mu$, and $\mu,\nu=x,y,z$. For more details, see Sec.~\ref{sec:idealpulse}, \ref{sec:avefinitepulse} for conditions 1, 2, Sec.~\ref{sec:avefinitepulse} for condition 3 and definition of the ``parity" of frame changes, Sec.~\ref{sec:angleerror} for condition 4 and definition of the ``chirality" of frame changes.
\label{tab:summary}}
\end{center}
\end{figure*}
\renewcommand{\figurename}{FIG.}
\setcounter{figure}{2}

\subsection{Decoupling Conditions for Ideal Pulses}
\label{sec:idealpulse}
Our goal here is to perform dynamical decoupling and suppress both disorder and interaction effects, by generating a pulse sequence with a vanishing $H_\text{avg} = 0$~\cite{slichter2013principles,mehring2012principles,levitt2001,rhim1971time,burum1979analysis,cory1990time,rose2018high}. Examining the above expressions in Eqs.~(\ref{eq:Havedis}-\ref{eq:HaveAntiSym}), we observe that there are two types of functional dependencies on $F_{\mu,k}$: the disorder [Eq.~(\ref{eq:Havedis})] and anti-symmetric spin-exchange [Eq.~(\ref{eq:HaveAntiSym})] Hamiltonians involve terms linear in $F_{\mu,k}$, while the Ising [Eq.~(\ref{eq:HaveIsing})] and symmetric spin-exchange [Eq.~(\ref{eq:HaveSym})] Hamiltonians involve terms quadratic in $F_{\mu,k}$.

The first type of contribution, which depends linearly on $F_{\mu,k}$, can be cancelled if $K_\mu$ = 0 for all $\mu=x,y,z$ axes [see Eqs.~(\ref{eq:Havedis}, \ref{eq:HaveAntiSym})], giving
\begin{align}
	\sum_{k=1}^n F_{\mu,k} \tau_k = 0, \quad \text{for every $\mu=1,2,3$}.
\label{eq:ideal_decoupling}
\end{align}
For equidistant pulse sequences where $\boldsymbol{\tau} = \tau {\bf I}_{1 \times n}$, the above condition can be further simplified to $ \sum_{k=1}^n F_{\mu,k} = 0$. This suggests that each row $\mu$ of the matrix ${\bf F}$ should have an equal number of positive and negative elements, such that their sum is 0, resulting in $H_\text{avg}^\text{dis} = H_\text{avg}^\text{A} = 0$. Physically, this corresponds to guaranteeing a spin-echo-type structure, in which each precession period around a positive axis is compensated by an equal precession period in the opposite direction. Applying this criterion to the sequences in Fig.~\ref{fig:fig2}, we see that as expected, the CPMG [Fig.~\ref{fig:fig2}(a,d)] and echo+WAHUHA [Fig.~\ref{fig:fig2}(c,f)] sequences cancel on-site disorder. The WAHUHA sequence, however, produces a residual on-site disorder term, also known as the chemical shift, given by $H_\text{avg}^\text{dis} = \sum_i \frac{h_i}{3} (S_i^x + S_i^y  + S_i^z)$ [Fig.~\ref{fig:fig2}(b,e)].

In contrast, the terms with quadratic dependence on $F_{\mu,k}$ cannot be fully suppressed in general, as the isotropic (Heisenberg) component of the interaction is invariant under global rotations~\cite{choi2017dynamical}, leading to $H_\text{avg}^\text{int} \neq 0$. However, it is still possible to fully symmetrize these interactions into a Heisenberg Hamiltonian, $H_\text{avg}^\text{int} = \frac{1}{3} \sum_{ij} (J_{ij}^I + 2J_{ij}^S) \vec{S}_i \cdot \vec{S}_j$, which is in fact sufficient to preserve spin coherence in many situations; In particular, globally polarized initial states that are typically prepared in experiments constitute an eigenstate of the Heisenberg interaction, and consequently do not dephase under the Heisenberg Hamiltonian. Such interaction symmetrization is satisfied when $L_x = L_y = L_z$, giving
\begin{align}
	\sum_{k=1}^n |F_{x,k}| \tau_k = \sum_{k=1}^n |F_{y,k}| \tau_k  = \sum_{k=1}^n |F_{z,k}| \tau_k.
\label{eq:ideal_interaction}
\end{align}
Again, for equidistant pulses, this condition is simplified to the statement that the sum $\sum_{k=1}^n |F_{\mu,k}|$ should be the same for each $\mu$. Based on this analysis, we verify that the CPMG sequence [Fig.~\ref{fig:fig2}(a,d)] does not symmetrize interactions, since it only employs $\pi$ pulses, while the two sequences that incorporate $\pi/2$ pulses to switch between all axes in the toggling frame [Fig.~\ref{fig:fig2}(b,c,e,f)] do indeed symmetrize the interaction Hamiltonian into the Heisenberg form. The spin-1/2 dipolar interaction with $J^I_{ij} = -2J^S_{ij}$ is a special case where the WAHUHA sequence fully cancels interactions to leading order, giving $H_\text{avg}^\text{int} =0$.

While we have focused on single-body and two-body interactions, the above analysis can be extended to interactions involving more spins. In particular, in Sec.~\ref{sec:multibody} and Appx.~\ref{suppsec:threebody}, we utilize results from unitary $t$-designs~\cite{dankert2009exact,webb2015clifford,zhu2017multiqubit} to prove that the conditions described above also guarantee decoupling of general \textit{three-body} interactions for polarized initial states.

\section{Robust Pulse Sequence Design}
\label{sec:imperfect}
For pulses of finite duration, on-site disorder and interaction effects acting during the pulses cause additional dynamics in the quantum system~\cite{rhim1974analysis,burum1979analysis,cory1990time}. In addition, the spin rotations can also suffer from experimental control errors, such as over- or under-rotations. Both of these imperfections can contribute to an error Hamiltonian $\delta H_\text{avg}$, which can be estimated to leading order using AHT as 
\begin{align}
\delta H_\text{avg} = \frac{t_p}{T} \sum_{k=1}^n \bar{H}^{(0)}_{P_k},
\label{eq:Himperfect}
\end{align}
where $t_p$ is the duration of a $\pi/2$ pulse and $\bar{H}^{(0)}_{P_k}$ is the zeroth-order average Hamiltonian acting during the $k$-th $\pi/2$ pulse building block.
Thus, the total leading-order effective Hamiltonian describing the driven spin dynamics is given by 
\begin{align}
	H_\text{eff} = H_\text{avg} + \delta H_\text{avg}.
\label{eq:Have}
\end{align}
The goal of robust Hamiltonian engineering is to suppress the error $\delta H_\text{avg} = 0$ by designing leading-order fault-tolerant, self-correcting pulse sequences.

\begin{figure}[!ht]
\begin{center}
\includegraphics[width=\columnwidth]{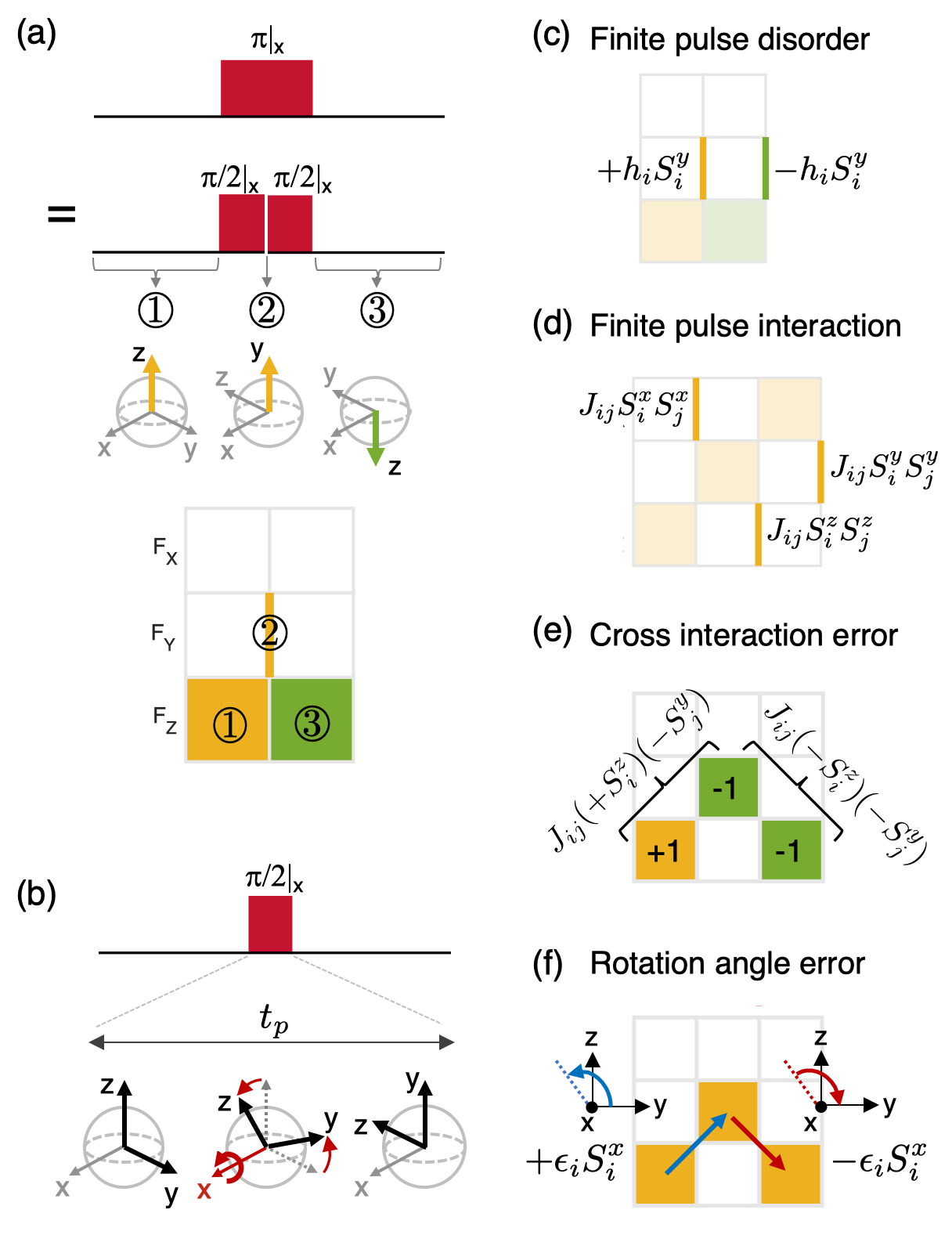}
\caption{{\bf Understanding control imperfections due to finite pulse duration and spin-manipulation error}. (a) $\pi$-pulse treatment. A $\pi$ pulse should be interpreted as two consecutive $\pi/2$ pulses with zero time separation. The intermediate frame after the first $\pi/2$ pulse is  represented as a thin line at the interface of the sequence matrix. (b) Finite pulse duration effect. Under a $\pi/2$ pulse, the spin frame smoothly rotates over the finite pulse duration $t_p$, which introduces errors in dynamic Hamiltonian engineering. (c)-(f) Leading-order error terms added to target effective Hamiltonian due to finite pulse duration and rotation angle errors; (c) On-site disorder Hamiltonians during the finite-duration pulse. To cancel them, the sequence matrix should contain an equal number of positive (yellow) and negative (green) intermediate frames for each axis. (d) Two-body Ising interactions during the finite-duration pulse. To suppress their effects, the sequence matrix should contain an equal number of intermediate frames for all three axes. (e) Two-body interaction cross-terms. To cancel them, the sequence matrix should contain an equal number of even and odd parities for every pair of neighboring frames between two axes; if the spin frame maintains (changes) its sign when switching to a different axis, the parity is defined as even (odd). (f) Spin-rotation angle error effects. To cancel them, the matrix should contain an equal number of positive and negative chiralities for every axis; the rotation axis and sign of the spin frame rotation in the toggling frame define positive/negative chirality along that axis.}
\label{fig:fig3}
\end{center}
\end{figure}

\subsection{Average Hamiltonian Analysis for Finite Pulse Duration}
\label{sec:avefinitepulse}
Turning to analyze finite pulse duration effects, we now provide an efficient method to understand and correct all pulse-related control errors. Here, the key insight is that our matrix representation ${\bf F}$ directly provides a simple way to obtain $\delta H_\text{avg}$ [Eq.~(\ref{eq:Himperfect})].
Intuitively, the form of $\bar{H}_{P_k}^{(0)}$ in $\delta H_\text{avg}$ is expected to be the average of the neighboring toggling-frame Hamiltonians, $\tilde{H}_{k}$ and $\tilde{H}_{k+1}$, since the finite-duration pulse $P_k$ smoothly changes the spin frame from the $k$-th to the $k+1$-th toggling frame [Fig.~\ref{fig:fig3}(b)]. However, as shown below, detailed calculations reveal nontrivial prefactors, as well as an additional interaction cross-term that can be expressed as a parity condition on neighboring matrix columns.

Since our pulse sequences are constructed out of $\pi/2$-pulse building blocks, we can analytically calculate $\bar{H}_{P_k}^{(0)}$ originating from the finite-duration pulse $P_k$ as
\begin{align}
\bar{H}^{(0)}_{P_k}=\frac{1}{t_p}\int_0^{t_p} (P_{k-1} \cdots P_1)^\dagger \tilde{H}_{P_k}(t) (P_{k-1} \cdots P_1) dt, 
\label{eq:finitepulse}
\end{align}
where $\{P_1,\cdots,P_{k-1}\}$ denotes the preceding $k-1$ rotations and the effect of the pulse $P_k$ is given by 
\begin{align}
	\tilde{H}_{P_k}(t) &= Q^\dagger_k(t) H_s Q_k(t).
\end{align}
Here, $Q_k(t) = \exp\qty[-i\sum_i \Omega_i t S_i^\nu]$ is the time-dependent unitary operator due to the pulse $P_k$ that globally rotates spins along the $\nu$-axis over the finite duration $0 \le t \le t_p$ and $\Omega_i$ is the Rabi frequency for a spin at site $i$, producing the $\pi/2$ rotation. For now, we assume no rotation angle errors: the treatment of them will be discussed in Sec.~\ref{sec:angleerror}.

Physically, the role of $\pi/2$ pulses is to smoothly interpolate the toggling-frame spin operator $\tilde{S}^z(\theta)=\cos\theta\tilde{S}_k^z+\sin\theta\tilde{S}_{k+1}^z$ during the finite pulse duration, where $\theta = \Omega t$. Using this to evaluate the integral of Eq.~(\ref{eq:finitepulse}), as detailed in Sec.~S1B~\cite{SM}, we obtain
\begin{align}
\bar{H}^{(0)}_{P_k}= &\frac{4}{\pi}\left[\left(\frac{\tilde{H}_k^{\text{dis}}+\tilde{H}_{k+1}^\text{dis}}{2}\right) + \left(\frac{\tilde{H}_k^\text{A}+\tilde{H}_{k+1}^\text{A}}{2}\right) \right]\nonumber\\
&+\left[\left(\frac{\tilde{H}_k^{I}+\tilde{H}_{k+1}^{I}}{2} \right)+\left(\frac{\tilde{H}_k^{S}+\tilde{H}_{k+1}^{S}}{2} \right) +\tilde{H}_{k,k+1}^C \right], 
\label{eq:Hpi2}
\end{align}
where $\tilde{H}_k^{\text{dis}}$, $\tilde{H}_k^{\text{I}}$, $\tilde{H}_k^{\text{S}}$ and $\tilde{H}_k^{\text{A}}$ are the disorder, Ising, symmetric and anti-symmetric spin-exchange interaction Hamiltonians at the $k$-th toggling frame, respectively, which can be obtained from replacing the original spin operators $S^\mu$ in $H_s$ [Eq.~(\ref{eq:genHam})] to the toggling-frame ones $\tilde{S}^\mu_k$, as shown in the individual terms in the summations of Eqs.~(\ref{eq:Havedis})-(\ref{eq:HaveAntiSym}). While most terms in Eq.~(\ref{eq:Hpi2}) are the weighted average of the neighboring toggling-frame Hamiltonians, consistent with the original intuition, there is an additional average Hamiltonian contribution $\tilde{H}_{k,k+1}^{\text{C}}$ resulting from two-body interaction cross-terms $\tilde{S}^z_{k}\tilde{S}^z_{k+1}$ acting during the pulse and given by
\begin{align}
	\tilde{H}_{k,k+1}^{\text{C}} &= \sum_{ij} J_{ij}^C \qty[\qty(\tilde{S}^z_k)_i \qty(\tilde{S}^z_{k+1})_j + \qty(\tilde{S}^z_{k+1})_i \qty(\tilde{S}^z_k)_j] \nonumber \\
						&= \sum_{ij, \mu\nu} C_{ij}^{\mu\nu} \sum_{k=1}^n \mathcal{P}_k^{\mu\nu} \label{eq:crossInt},
\end{align}
where $C_{ij}^{\mu\nu} = \sum_{ij} J^C_{ij} (S^\mu_i S^\nu_j + S^\nu_i S^\mu_j)$ is the cross-interaction operator with $J_{ij}^C = \frac{1}{\pi}(J^I_{ij}-J^S_{ij})$ (see Sec.~S1B~\cite{SM}) and $\mathcal{P}_k^{\mu\nu}$ defines the ``parity" of neighboring $k$-th and $k+1$-th frames, given as
\begin{align}
    \mathcal{P}_k^{\mu\nu} = F_{\mu,k} F_{\nu,k+1} + F_{\nu,k}F_{\mu,k+1}.
\end{align}
To cancel the interaction cross-terms, the parity should vanish when summed over one Floquet period for each pair ($\mu,\nu$): $\sum_{k=1}^n \mathcal{P}_k^{\mu\nu} = 0$ [Fig.~\ref{fig:fig3}(e)]. Intuitively, the parity can be understood as checking whether the signs of neighboring frames are the same ($\mathcal{P}_k^{\mu\nu} =+1$, even parity) or different ($\mathcal{P}_k^{\mu\nu} =-1$, odd parity).

Taking into account the distinct weighting factors of $4/\pi$ and 1 for the different interaction types as well as the additional interaction cross-term, as identified in Eq.~(\ref{eq:Hpi2}), the effective Hamiltonian in the presence of finite pulse duration $t_p$, $H_\text{eff} = H_\text{avg} + \delta H_\text{avg}$, becomes
\begin{align}
	H_\text{eff} = \frac{1}{T} &\left[\sum_{k=1}^n \qty(\tau_k + \frac{4}{\pi}t_p ) \qty(\tilde{H}^\text{dis}_k + \tilde{H}^\text{A}_k)\right. \nonumber \\
						&+ \left.\sum_{k=1}^n \qty(\tau_k + t_p) \qty(\tilde{H}^\text{I}_k + \tilde{H}^\text{S}_k)+ \sum_{k=1}^n t_p H^\text{C}_{k,k+1}\right],
\label{eq:Havrg}
\end{align}
where the base pulse sequence length, $T = (\sum_{k=1}^n \tau_k) + n t_p$, includes the total length of $n$ $\pi/2$ pulses. As described above in the discussion following Eq.~(\ref{eq:Hpi2}), each of the terms in the toggling-frame Hamiltonian, $\tilde{H}^\text{dis}$, $\tilde{H}^\text{A}$, $\tilde{H}^\text{I}$, $\tilde{H}^\text{S}$, and $\tilde{H}^\text{C}$, can be readily computed using our sequence representation.

\subsection{Analysis of Rotation Angle Error}
\label{sec:angleerror}
We now analyze the effects of rotation angle errors in control pulses, resulting from imperfect and inhomogeneous global spin manipulation. At first glance, this may seem challenging, since the average Hamiltonian corresponding to a rotation angle error around a given axis in the lab frame depends on the transformations by previous pulses. However, we find that there is a simple intuition for their average Hamiltonian contribution in the \textit{toggling frame}, whereby an imperfect rotation around the $+\hat{\mu}$ direction (positive chirality) in the toggling frame can be compensated by another rotation around the $-\hat{\mu}$ direction (negative chirality). Moreover, the rotation axis in the toggling frame can be readily described using our frame matrix $\mathbf{F}$, allowing concise conditions based on rotation chirality to achieve self-correction of rotation angle errors in a pulse sequence.

More specifically, the rotation axis in the $k$-th toggling frame, $\vec{\beta}_k=\sum_\mu \beta_{\mu,k}\hat{e}_\mu$, can be obtained by taking the cross product of the frame vectors before and after the pulse
\begin{align}
\vec{\beta}_k=\vec{F}_{k+1}\times\vec{F}_{k},
\label{eq:betachirality}
\end{align}
where $\vec{F}_k =\sum_\mu F_{\mu,k} \hat{e}_\mu$. Physically, this cross product structure can be thought of as characterizing the chirality of the toggling-frame rotation from $\vec{F}_k$ to $\vec{F}_{k+1}$. As derived in more detail in Sec.~S1C~\cite{SM}, the average Hamiltonian contribution corresponding to the rotation angle error is then given by
\begin{align}
\delta H_\text{avg}^\text{rot}=\frac{1}{T}\sum_{i,\mu} \epsilon_i \qty( \sum_{k=1}^n \beta_{\mu,k} S_i^\mu),
\end{align}
where $\epsilon_i$ is the static rotation-angle deviation from the target $\pi/2$ angle for a spin at site $i$. This allows us to identify the cancellation condition for rotation angle errors, $\delta H_\text{avg}^\text{rot}=0$, as 
\begin{align}
\sum_k \beta_{\mu,k} = 0 \quad \text{for each $\mu = x,y,z$}
\label{eq:anglecorr}
\end{align}
which corresponds to condition 4 in Tab.~\ref{tab:summary}.

\subsection{Decoupling Conditions for Finite Duration Pulses}
\label{sec:analyzefinitepulse}
By incorporating the dominant effects arising from finite pulse durations, we have obtained a more realistic form of the effective Hamiltonian. In the leading-order average Hamiltonian, we found that the finite pulse duration simply introduces corrections to the effective free evolution intervals in Eqs.~(\ref{eq:ideal_decoupling},\ref{eq:ideal_interaction}), plus additional terms that are well-described by the parity and chirality associated with each toggling frame change. Combining these, we arrive at the conditions to achieve robust dynamical decoupling for a polarized initial state, as summarized in Tab.~\ref{tab:summary}.

Remarkably, we have found that our matrix representation ${\bf F}$ provides a systematic treatment of all such imperfections in a simple, pictorial fashion.
As an example, for the common case of equidistant pulses, condition 1 in Tab.~\ref{tab:summary} is satisfied when there is an equal number of yellow ($F_{\mu,k}$ = $+1$) and green ($F_{\mu,k}$ = $-1$) squares (lines) in each row, while condition 2 is satisfied when different rows have an equal number of squares, and an equal number of lines. Moreover, in Sec.~\ref{sec:multibody} and Appx.~\ref{suppsec:threebody}, we further show that our representations and methods can be applicable to more complex three-body interactions, even for finite pulse durations, highlighting the broad applicability of our sequence design framework.

Additional effects such as waveform transients and pulse shape imperfections can also be analyzed by a similar approach~\cite{rhim1974analysis} (see for example Sec.~S1D~\cite{SM} for the treatment of rotation axis imperfections).

\section{Extensions to Higher-order Average Hamiltonians and Multi-Body Interactions}
\label{sec:HighOrder}
\subsection{Suppression of Higher Order Effects}
While our analysis thus far has focused on the zeroth-order average Hamiltonian, we can incorporate higher-order effects by considering the full Magnus expansion [Eq.~(\ref{eq:Magnus})] to engineer effective Hamiltonians with higher accuracy. More specifically, our frame matrix representation allows us to readily evaluate the higher-order expansion terms, which consist of commutators between Hamiltonians at different times in the toggling frame (see Appx.~\ref{sec:AHT}). For example, the first-order contribution [Eq.~(\ref{eq:Magnus1})] for a periodic pulse sequence, including finite pulse effects, can be expressed as 
\begin{align}
	\bar{H}^{(1)}&=-\frac{i}{2T} \sum_{k=1}^n \sum_{l=1}^{k} \qty[\Theta_l,\Theta_k]+O(t_p^2),
	\label{eq:H1term}
\end{align}
where $n$ is the total number of evolution intervals and $\Theta_{l,k}$ are the time-weighted Hamiltonians in the $l$-th and $k$-th toggling frames, respectively, given as (including finite-pulse duration effects)
\begin{align}
	\Theta_k &= \tau_k (\tilde{H}^\text{dis}_k + \tilde{H}^\text{A}_k + \tilde{H}^\text{I}_k + \tilde{H}^\text{S}_k) \nonumber \\
			&+ t_p  \left[\frac{4}{\pi}(\tilde{H}^\text{dis}_k + \tilde{H}^\text{A}_k) + \tilde{H}^\text{I}_k + \tilde{H}^\text{S}_k+ H^\text{C}_{k,k+1}\right],
\end{align}
and the $O(t_p^2)$ term coming from commutations of a finite pulse Hamiltonian with itself will typically be small (there are $n$ such terms, compared to $n^2/2$ terms for the $\Theta$ terms, and $t_p$ is typically small). Recall from Eq.~(\ref{eq:Havrg}) that the zeroth-order Hamiltonian is simply $\bar{H}^{(0)} = \frac{1}{T} \sum_{k=1}^n \Theta_k$. Crucially, note that all $\Theta_{k=\{1,\cdots,n\}}$ are easily numerically computable with our matrix representation that specifies the toggling-frame evolution of the $S^z$ operator [Eqs.~(\ref{eq:Havedis}-\ref{eq:HaveAntiSym})]. This enables us to readily evaluate the contribution from the first-order term [Eq.~(\ref{eq:H1term})], which will result in algebraic conditions that involve second-order polynomials in $F_{\mu,k}$ and $|F_{\mu,k}|$. Analyzing the second-order or even higher-order terms is also straightforward as they can be obtained in a very similar fashion via recursive computation of nested commutators involving $\Theta_k$'s at different frames.

As an explicit example, the first-order term for the echo+WAHUHA sequence [Fig.~\ref{fig:fig2}(c,f)] can be analytically derived to be (assuming uniform $\tau_k$)
\begin{align}
\bar{H}^{(1)}\approx&\sum_{ij} \frac{1}{6}   \qty(2J^I_{ij}+3J^S_{ij})   \qty(h_i-h_j)\qty(S_i^xS_j^y-S_i^yS_j^x)\nonumber\\+&\frac{1}{6}\qty(h_i-h_j)J^S_{ij}\qty(S_i^yS_j^z-S_i^zS_j^y)\nonumber\\-&\frac{1}{2}J^S_{ij}\qty(h_iS_i^xS_j^z-h_jS_i^zS_j^x)\nonumber\\+&\frac{1}{6}\qty(2J^I_{ij}+J^S_{ij})\qty(h_jS_i^xS_j^z-h_iS_i^zS_j^x),
\end{align}
where we have simplified the expression by dropping anti-symmetric exchange interactions that are typically not present, and assuming $t_p\ll\tau_k$.

In addition to explicitly evaluating the higher-order expansion terms, one can also use various heuristics to suppress higher-order terms and enhance the accuracy of Hamiltonian engineering. Developed primarily in the NMR community, there are several known approaches, such as reflection-symmetric pulse arrangements~\cite{mansfield1971symmetrized,burum1979analysis,cory1990time,li2007generating} and concatenated sequence symmetrization~\cite{khodjasteh2005fault,souza2011robust,wang2012comparison,farfurnik2015optimizing}, to suppress higher-order contributions in driven spin dynamics. In the following, we discuss how these techniques can also be naturally incorporated into our sequence design framework.

\textit{Reflection symmetry}: When Hamiltonians in the toggling frame respect reflection symmetry~\cite{mansfield1971symmetrized}, that is, $\tilde{H}_{n+1-k} = \tilde{H}_k$, all odd-order terms in the Magnus expansion vanish, $\bar{H}^{(2l-1)}$ = 0 with integer $l$ [Eq.~(\ref{eq:Magnus})]. In our framework, this imposes an additional condition on the sequence ${\bf F}$; Generically however, any sequence can be extended into a pulse sequence that respects reflection symmetry simply by doubling the length of the frame matrix and filling the second half with its own mirror image in time, taking care of pulse imperfections at the central interface. As an example, we can apply the reflection symmetry to the echo+WAHUHA sequence [Fig.~\ref{fig:fig2}(c,f)] to cancel higher-order effects, as shown in Fig.~\ref{fig:fig4}(a,b). We note, however, that for certain applications such as quantum sensing, one needs to take additional care when performing such symmetrizations, since reflection symmetry (similar to a time-reversal operation) may accidentally cancel the desired sensing field contributions [Sec.~\ref{sec:sensing}].

\textit{Concatenated sequence symmetrization}: One way to understand and engineer higher-order Hamiltonian engineering properties of a sequence is to decompose it into smaller building blocks. A few techniques developed along these lines include pulse-cycle decoupling~\cite{burum1979analysis} and concatenated symmetrization schemes~\cite{khodjasteh2005fault,souza2011robust,wang2012comparison,farfurnik2015optimizing}, where a long pulse sequence is constructed from the repetition of short pulse sequences, symmetrized in a systematic pattern to suppress higher-order effects. Our method can facilitate the robust implementation of such concatenation schemes by providing both an intuitive visualization and precise algebraic conditions to analyze the error robustness of concatenated pulse sequences.

\textit{Second averaging}: The technique of second-averaging has been developed in NMR to suppress dominant error terms in $\delta H_\text{avg}$ that do not commute with the leading contribution in $H_\text{avg}$~\cite{haeberlen1971resonance,cory1990multiple,cory1996distortions}. Such methods can be readily incorporated in our framework by alternating the rotation axes of control pulses periodically every Floquet cycle, or by using off-resonant driving.

\subsection{Enhanced Numerical Search of Pulse Sequences}
\label{sec:numerics}
Our formalism not only enables efficient pen-and-paper pulse sequence design and provides important analytical insights, but can also greatly enhance the numerical search of pulse sequences. More specifically, the concise decoupling rules we have derived above provide a rapid means to narrow the search space down to pulse sequences that may have good performance, as a starting point for in-depth numerical simulations that capture the full dynamics of the system to all orders.

To illustrate this with a concrete example, we consider pulse sequences with 12 free evolution intervals that aim to efficiently decouple the effects of interactions and disorder. An exhaustive search of just such pulse sequences ignoring finite pulse duration effects would already require an enumeration of $6^{12} \approx 10^{9}$ possibilities (6 possible configurations for each toggling frame, i.e., $F_{\mu,k} = \pm1$ for $\mu=x,y,z$), a prohibitively large number for numerical simulations. However, the application of our disorder- and interaction-decoupling rules on the generation of sequences can significantly narrow down the search space. In addition, depending on the target application, more constraints in the form of algebraic rules can be simultaneously applied to further reduce the size of the sequence space. For example, for efficient AC-field sensing we can impose a fast spin-echo structure whereby the signs of toggling frames are periodically flipped over the shortest possible period 2$\tau$ while maintaining a synchronized phase relation between different axes (see Fig.~\ref{fig:fig6}(b) as an example). Such a phase-locked, fast-echo structure acts as a bandwidth filter centered at a target frequency (see Sec.~\ref{sec:sensing} for detailed discussions). This allows us to find a total of 14,080 sequences that can be sorted out into four different categories according to their error robustness: Class I satisfies all decoupling rules (448 sequences), Class II does not fully decouple interaction cross-terms [violation of Condition 3 in Tab.~\ref{tab:summary}], Class III does not fully suppress rotation angle errors [violation of Condition 4 in Tab.~\ref{tab:summary}], and Class IV does not suppress interaction cross terms and rotation angle errors [violation of both Conditions 3, 4 in Tab.~\ref{tab:summary}].

\begin{figure}[!ht]
\begin{center}
\includegraphics[width=\columnwidth]{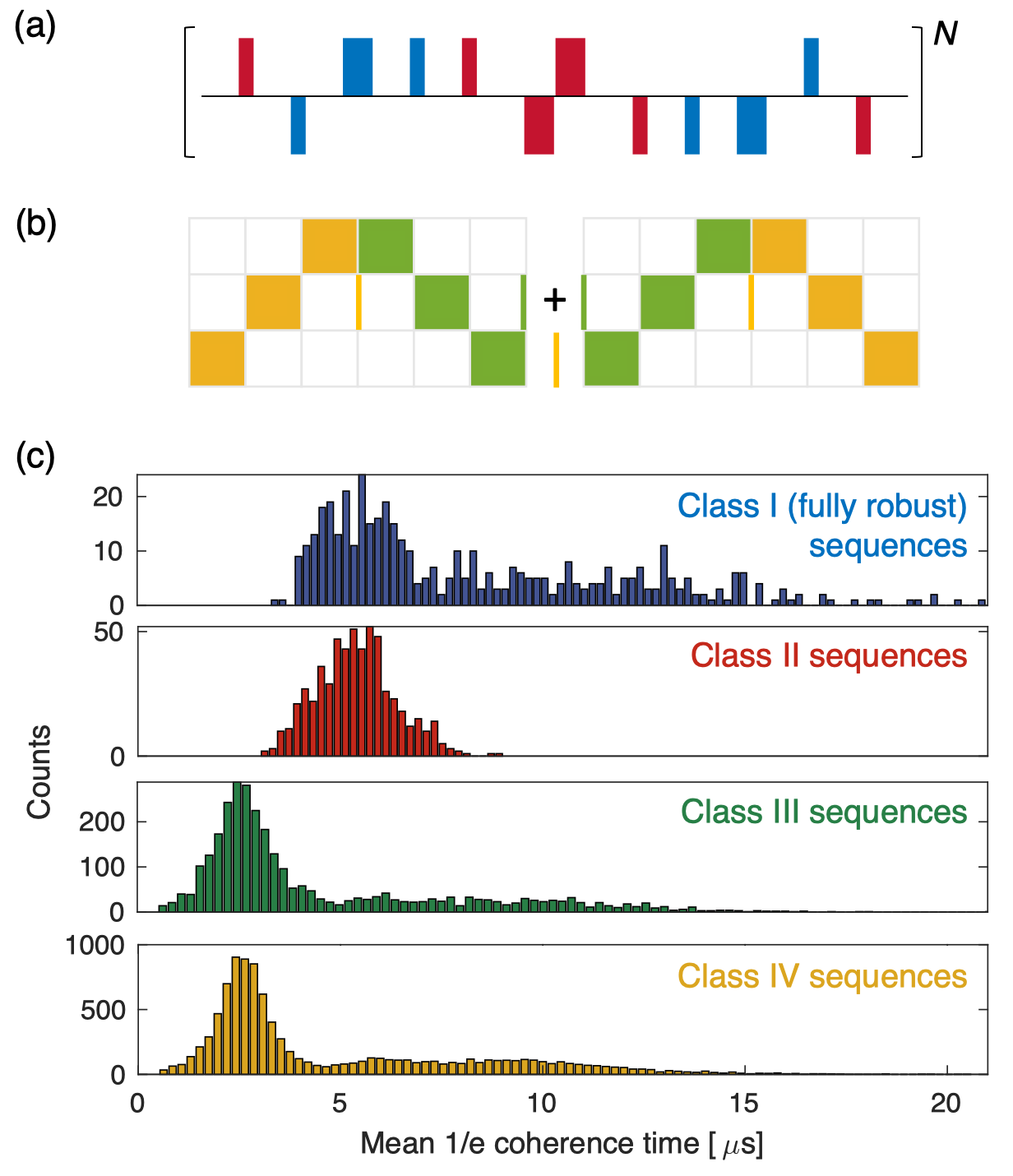}
\caption{{\bf Sequence symmetrization for higher-order suppression and numerical validation of the fault-tolerant decoupling rules}. (a) Symmetrized echo+WAHUHA sequence [Fig.~\ref{fig:fig2}(c)] and (b) its matrix-based sequence representation. The $\pi/2$ and $\pi$ pulses are depicted in the same way as in Fig.~\ref{fig:fig2}. In (b), the intermediate frames are explicitly shown to incorporate finite pulse effects. The middle yellow bar at the interface introduces a small, residual disorder and interaction in the zeroth-order average Hamiltonian, which can be compensated by slightly adjusting free evolution periods of neighboring toggling frames. Due to the reflection symmetry with respect to the center of the sequence, all odd-order expansion terms vanish. (c) Histograms of $1/e$ coherence decay times for classes of pulse sequences exhibiting different degrees of robustness. The sequences are generated by respecting the decoupling rules in Tab.~\ref{tab:summary}, which help to efficiently reduce the size of the sequence search space. For sequence evaluation, we performed exact diagonalization studies for a disordered, interacting 8-spin system (see text for simulation details). The different sequences are classified into four distinct categories, Class I (blue), II (red), III (green) and IV(yellow), according to their robustness to control imperfections (see text for class definitions).}
\label{fig:fig4}
\end{center}
\end{figure}

To evaluate the performance of the sequences, we numerically solve the exact Floquet spin dynamics for a disordered, interacting 8-spin system and monitor global spin polarization as a function of time.We choose Gaussian random on-site disorder ($\sigma_W=(2\pi)4.0$ MHz) and uniform random interactions ($J=(2\pi)0.2$ MHz), with pulse spacing $\tau=25$ ns and pulse duration $t_p=10$ ns, and extract the $1/e$ coherence decay times $T_2$ averaged over $\hat{x}, \hat{y}, \hat{z}$ initial states (with the average performed over decay rates). As shown in Fig.~\ref{fig:fig4}(c), Class I pulse sequences, satisfying all decoupling rules, perform considerably better than the other classes. In particular, we find that the top 10 sequences with longest coherence times all consistently belong to Class I. Note that the numerically-optimized sequences exhibit a broad distribution of coherence times due to different amounts of contributions from higher-order terms in the Magnus expansion. Indeed, using Eq.~(\ref{eq:H1term}), we explicitly verify in Sec.~S1F~\cite{SM} that the resulting coherence decay is strongly correlated with the first order contribution. These results further confirm that the analytical insights provided by our formalism can substantially improve the numerical search efficiency for optimal pulse sequences, allowing fast numerical optimization that can capture effects from all orders.

\subsection{Extensions to Multi-Body Interactions}
\label{sec:multibody}
Our discussion thus far has focused on the case of one- and two-body interactions. Interestingly, our versatile formalism can also be applied to more complex scenarios, leading us to a new set of rules that allow for the implementation of robust protocols in the presence of three-body interactions. In particular, we show via a neat connection to unitary $t$-designs~\cite{divincenzo2002quantum,dur2005standard,emerson2003pseudo,collins2006integration,gross2007evenly,ambainis2007quantum,dankert2009exact,webb2015clifford,zhu2017multiqubit} that (i) in the limit of ideal pulses, the decoupling conditions described in Sec.~\ref{sec:idealpulse} are also sufficient to fully suppress dynamics under any secular three-body interaction for a polarized initial state, and (ii) we can extend the formalism that accounts for finite pulse duration effects to the case of three-body interactions, leading to new decoupling conditions beyond those discussed in Sec.~\ref{sec:imperfect}. Such interactions are important building blocks of exotic topological phenomena~\cite{moore1991nonabelions,fradkin1998chern,moessner2001resonating,levin2005string}, and have been proposed to be realized in cold molecules~\cite{buchler2007three} and superconducting qubits~\cite{mezzacapo2014many,chancellor2017circuit}. We sketch the main ideas of the derivation here, and detailed proofs can be found in Appendix.~\ref{suppsec:threebody}.

First, let us consider the case of perfect, infinitely short pulses. We will show, via connections to unitary $t$ designs, that under the above decoupling conditions, a polarized initial state will be an eigenstate of the resulting symmetrized Hamiltonian. A unitary $t$-design is a set of unitary operators $\{U_k\}$, such that
\begin{align}
   \frac{1}{K}\sum_{k=1}^K (U_k^\dagger)^{\otimes N}\mathcal{O}U_k^{\otimes N}=\int_{\mathcal{U}(2)}dU (U^\dagger)^{\otimes N}\mathcal{O}U^{\otimes N}\triangleeq \mathcal{O}_U. 
   \label{eq:unitarydesign}
\end{align}
Here, $\mathcal{U}(2)$ is the unitary group of dimension 2, used to describe two-level systems, $\mathcal{O}$ is an $N$-body operator with $N \leq t$ and $\mathcal{O}_U$ is the corresponding averaged observable. Intuitively, this expression means that for observables up to order $t$, the effect of averaging over the finite set of unitary operators $\{U_k\}$ is equivalent to averaging over all unitaries of dimension 2.

The symmetrizing properties of the right-hand-side of Eq.~(\ref{eq:unitarydesign}), where the average is taken over all elements of the unitary group over the Haar measure, imply that $\mathcal{O}_U$ must only contain terms proportional to elements of the symmetric group $S_t$ of order $t$~\cite{collins2006integration}. This is because all other terms will be transformed and symmetrized out by the average, but elements of the symmetric group, which only permute the labels of the states, will be invariant, as the unitary operator $U^{\otimes N}$ conjugates all spins identically.

It is known that the Clifford group forms a unitary 3-design~\cite{webb2015clifford,zhu2017multiqubit}. Combined with the fact that for interactions under the secular approximation, averaging over the Clifford group is equivalent to averaging over the six axis directions (see Appx.~\ref{suppsec:threebody}), this implies that for any sequence that satisfies the above decoupling rules, all interactions involving three particles or fewer will be symmetrized into a form that only contains terms proportional to elements of the symmetric group. Any initial state with all spins polarized in the same direction will then be an eigenstate of this symmetrized interaction, since this state is invariant under any permutation of the elements. Correspondingly, a polarized initial state does not experience decoherence under this interaction.

As a nontrivial example of this result, let us consider the interaction $H_\text{int}=J \sum_{ijk} (S_i^xS_j^yS_k^z-S_i^yS_j^xS_k^z)$. The symmetrized Hamiltonian can be calculated to be $\overline{H}_\text{int}=\frac{J}{3}\sum_{ijk} \sum_{\mu\nu\sigma}\epsilon_{\mu\nu\sigma}S_i^\mu S_j^\nu S_k^\sigma$, where $\epsilon_{\mu\nu\sigma}$ is the Levi-Civita symbol. One can explicitly verify that any globally polarized initial state is an eigenstate of the symmetrized Hamiltonian with eigenvalue 0.

In fact, we can also extend this analysis to the case of finite pulse durations by expanding the Hamiltonian as a polynomial in $F_{\mu,k}$ and examining how different possible terms transform. As described in Appx.~\ref{suppsec:threebody}, this gives rise to new decoupling conditions in the three-body case, as a generalization of the interaction cross-term decoupling condition [Condition 3 in Tab.~\ref{tab:summary}].

\section{Application: Dynamical Decoupling}
\label{sec:decouple}

\begin{figure}[h!]
\begin{center}
\includegraphics[width=\columnwidth]{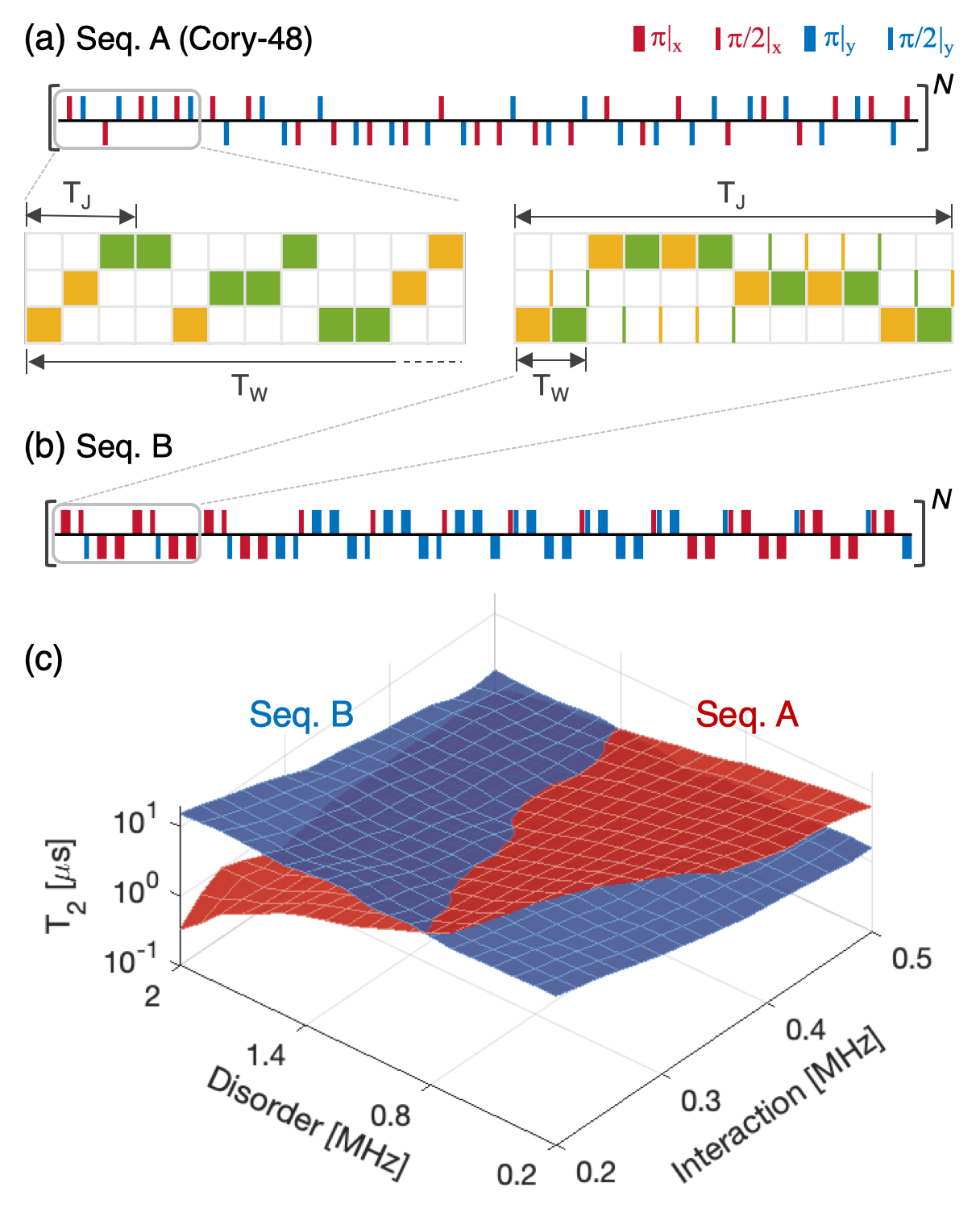}
\caption{{\bf System-targeted dynamical decoupling}. (a,b) Sequence details and their matrix representations. The inset illustrates the pulse legend used to denote $\pi/2$ and $\pi$ pulses. Decoupling  sequences can be characterized by their different timescales of disorder and interaction decoupling, denoted as $T_W$ and $T_J$, respectively. In (a), Seq. A (Cory-48) is designed for interaction-dominated systems, where axis permutation is performed on the fastest timescale to prioritize interaction symmetrization over disorder cancellation ($T_W \gg T_J$). In (b), Seq. B is designed for disorder-dominated systems where echo-like operations are performed on the fastest timescale to prioritize disorder suppression over interaction symmetrization ($T_W \ll T_J$). (c) Numerical simulation results of the dynamical decoupling performance of Seq.~A and Seq.~B for a wide range of disorder $W$ and interaction $J$ strengths. The figure of merit for the comparison is the coherence time $T_2$, extracted as the $1/e$ decay time of initially polarized spin states. More specifically, we perform an exact diagonalization simulation of a disordered, interacting ensemble of 12 spins. In the simulations, we initialize all spins along the $\hat{x}$-axis, periodically drive them with the pulse sequences, and measure the spin decoherence profile at stroboscopic times $t=NT$. Each parameter set is averaged over 100 disorder realizations and the averaged profile is used to identify $T_2$. The $\pi$ pulse duration and spacing are chosen to be $t_p = 25$~ns and $\tau=20$~ns, respectively. We have verified that choosing different $t_p$ values does not qualitatively change the results.}
\label{fig:fig5}
\end{center}
\end{figure}

\subsection{System-Targeted Dynamical Decoupling}
The goal of dynamical decoupling is to extend the coherence time by cancelling the effects of disorder and interactions, and by suppressing pulse imperfections in a robust fashion. In the field of NMR, a number of decoupling methods have been developed, such as multiple-pulse sequences~\cite{hahn1950spin,carr1954effects,meiboom1958modified,gullion1990new,viola1999dynamical,khodjasteh2005fault,viola2005random,uhrig2007keeping,khodjasteh2009dynamical,biercuk2009optimized,du2009preserving,alvarez2010performance,west2010near,de2010universal,ryan2010robust,suter2016,burum1979analysis,naydenov2011dynamical,rhim1971time,drobny1978fourier,burum1979analysis,shaka1983improved,baum1985multiple,shaka1988iterative,tycko1990zero,cory1990time,lee1995efficient,hohwy1999fivefold,carravetta2000symmetry,slichter2013principles,mehring2012principles,levitt2001,sorensen1984product}, magic echoes~\cite{takegoshi1985magic,rose2018high}, frequency- and phase-modulated continuous driving~\cite{lee1965nuclear,vinogradov1999high}, and numerically-optimized control schemes~\cite{iwamiya1993application,sakellariou2000homonuclear}.

However, these sequences are optimized for interaction-dominated dipolar-interacting spin systems only, making it difficult to extend them to other Hamiltonians exhibiting different energy scales and interaction forms. For example, electronic spin ensembles typically display strong on-site disorder with weak interactions~\cite{kucsko2018critical,choi2017depolarization}, such that a naive application of the NMR pulse sequences performs poorly (see Sec.~\ref{sec:experiment}). In the following, we show that our framework allows the design of system-targeted dynamical decoupling sequences that tackle the dominant effects on a faster timescale to achieve better performance.

As an example, for disorder-dominated systems, disorder cancellation needs to be prioritized and performed on a shorter timescale compared to interaction symmetrization and control error suppression. 
Our representation directly reveals the individual decoupling timescales $T_W$ and $T_J$ for disorder and interactions, respectively (illustrated in Fig.~\ref{fig:fig5}(a)): They can be quantified as the minimum length of toggling-frame time evolution that fulfills their respective decoupling conditions [Condition 1,2].
With $W$ and $J$ describing the characteristic disorder and interaction scales of the driven system, in the disorder-dominated case ($W \gg J$), we thus require $T_W \ll T_J$ to reduce the magnitude of higher-order contributions to $H_\text{eff}$~\cite{cory1991new,burum1979analysis,cory1990time} and maximize the leading-order approximation accuracy.
If the control error magnitude $\epsilon$ is comparable to $W,J$, then chirality cancellation [Condition 4] associated with pulse imperfections should also be performed at a relatively fast rate to suppress higher-order errors.

To illustrate the importance of system-targeted design, we provide two periodic pulse sequences (Seq.~A,~B) both designed to robustly decouple disorder and interactions, but with Seq.~A(B) better suited for systems characterized by stronger interactions(disorder). For Seq.~A, we adopt the Cory-48 sequence~\cite{cory1990time} developed for nuclear spin systems, where spin-spin interactions dominate over disorder. Indeed, as shown in Fig.~\ref{fig:fig5}(a), Seq.~A symmetrizes interactions very rapidly and also cancels on-site disorder, but on a much slower timescale ($T_J \ll T_W$). It is also robust against leading-order imperfections resulting from finite pulse durations, and suppresses certain higher-order effects~\cite{cory1990time}. For comparison, we design a new sequence, Seq.~B in Fig.~\ref{fig:fig5}(b), based on the conditions in Tab.~\ref{tab:summary}, to make the sequence operate better in the opposite, disorder-dominated regime. Specifically, Seq.~B incorporates frequent $\pi$ pulses to echo out disorder on a rapid timescale while symmetrizing interactions on a slower timescale ($T_W \ll T_J$). We emphasize that Seq.~B incorporates both $\pi$ pulses and composite $\pi/2$ pulses when switching between toggling frames [Fig.~\ref{fig:fig5}(b)], to accomplish fast spin-echo operations and retain robustness to control imperfections, and thus lies beyond the design capabilities of previous approaches.

Given these design considerations, we expect Seq.~A to perform better in the regime of large interaction strengths (e.g. for NMR), and Seq.~B to perform better for disorder-dominated systems (e.g. for electronic spin ensembles). From numerical simulations, as shown in Fig.~\ref{fig:fig5}(c), we indeed see a crossover in performance as the disorder and interaction strengths are tuned in the system within the range $0.009 < W(\tau+t_p) < 0.09$ and $0.009 < J(\tau+t_p) < 0.023$. At small disorder values, Seq.~A shows a longer coherence time $T_2$ than Seq.~B. However, as we increase the disorder strength, we observe a crossover beyond which Seq.~B outperforms Seq.~A. Overall, Seq.~B shows a stable performance within the range of parameters studied, while Seq.~A shows a strong susceptibility to disorder. This example illustrates how our formalism enables the systematic design of pulse sequences adapted to the dominant energy scales of different systems.

\subsection{Shortest Sequence for Robust Dynamical Decoupling}
\label{sec:Shortest}
The algebraic conditions introduced in Tab.~\ref{tab:summary} greatly simplify the design procedure, thereby allowing one to not only design pulse sequences that are robust against certain imperfections, but also guarantee via analytical arguments the shortest sequence length to achieve a set of target Hamiltonian engineering requirements.

The conditions to cancel disorder [Condition 1] and symmetrize interactions [Condition 2] at the average Hamiltonian level require an equal number of $\pm1$ frames along each axis, resulting in at least 6 distinct free evolution intervals when neglecting pulse imperfections, which implies that the echo+WAHUHA sequence in Fig.~\ref{fig:fig2}(c,f) has shortest length for ideal pulses. When incorporating finite pulse durations and rotation angle errors, the conclusion may be less obvious; however, using the algebraic conditions in Tab.~\ref{tab:summary}, we find that the following pulse sequence, consisting of 6 free evolution periods of duration $\tau$ connected by {\it composite} $\pi/2$ pulses, satisfies all leading-order decoupling requirements:
\begin{align}
\begin{pmatrix}
{\bf F} \\
\boldsymbol{\tau}
\end{pmatrix}_\text{Opt-6$\tau$}
=\left(\begin{array}{ccccccccccccc}
0 & 0 & 1 & 0 & 0 & -1 & 0 & 0 & -1 & 0 & 0 & 1 \\
0 & 1 & 0 & 0 & -1 & 0 & 0 & 1 & 0 & 0 & -1 & 0 \\
1 & 0 & 0 & -1 & 0 & 0 & -1 & 0 & 0 & 1 & 0 & 0\\
\tau & \tau & 0 & 0 & \tau & 0 &\tau & 0 & \tau & 0 & 0 & \tau\\
\end{array}\right) \nonumber.
\end{align}
To the best of our knowledge, this is the first pulse sequence that decouples all leading-order imperfections and achieves pure Heisenberg interactions with only 6 free evolution intervals, illustrating the power of our formalism. We note that the minimum achievable length of the pulse sequence may be modified by experimental considerations: for example, it may be challenging to apply pulses with different phases in close succession due to finite transient times of the experimental apparatus, and in such cases we can show that the minimum pulse length increases to 12 pulses, see Appx.~\ref{suppsec:optimal} for details.

\section{Application: Quantum Sensing with Interacting Spin Ensembles}
\label{sec:sensing}
\begin{figure*}
\begin{center}
\includegraphics[width=2\columnwidth]{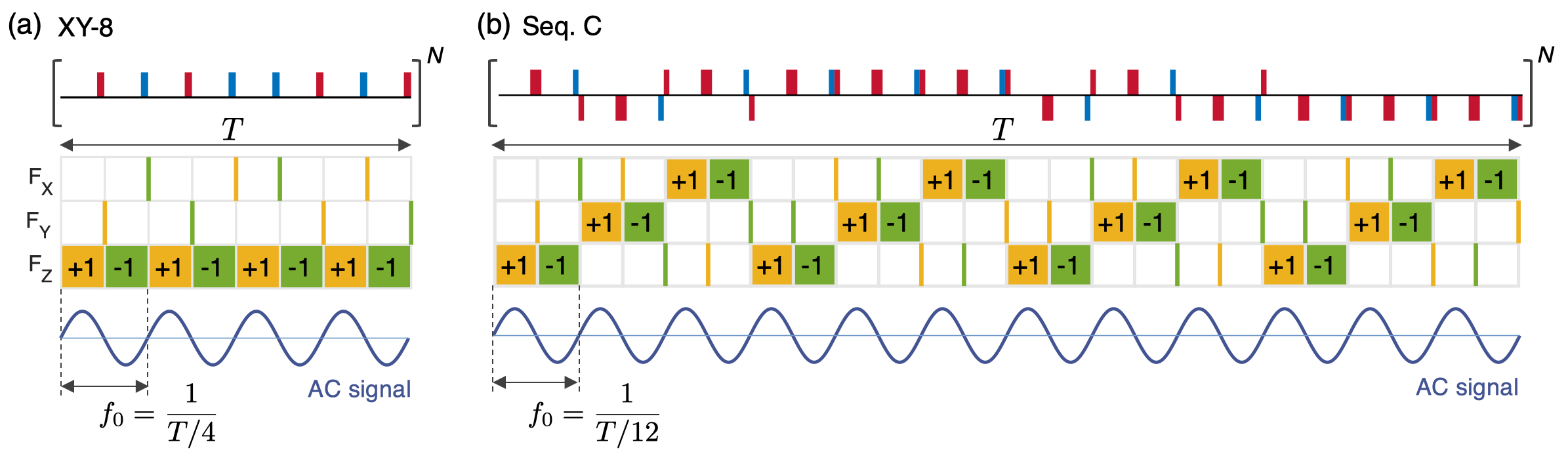}
\caption{{\bf Optimal AC signal sensing with robust pulse sequences}. (a,b) Sequence details for (a) the conventional XY-8 sequence optimized for non-interacting systems and (b) the robust sensing sequence, Seq.~C, optimized for interacting systems. The matrix representations, ${\bf F} = [\vec{F}_x;\vec{F}_y;\vec{F}_z]$, for each sequence are shown with resonant AC signals (blue curves) at the frequency $f_0$ to be detected. The sequence length, $T$, includes both free evolution periods and finite pulse durations. The $\pi/2$ and $\pi$ pulses are depicted in the same way as in the pulse legend of Fig.~\ref{fig:fig5}. In (b), composite pulses consisting of two $\pi/2$ pulses are extensively used to preserve sequence robustness while performing AC-field sensing (see Sec.~\ref{sec:senseopt} for a detailed discussion).}
\label{fig:fig6}
\end{center}
\end{figure*}

Quantum sensing presents additional challenges beyond the simple decoupling of the effects that cause decoherence. Here, in addition to \textit{decoupling} disorder and interactions to extend coherence time, one also needs to \textit{recouple} the target signal to perform effective sensing. While there has been extensive research for quantum sensing with non-interacting systems (e.g. the XY-8 sequence~\cite{naydenov2011dynamical,de2011single}), there are only a limited number of such demonstrations for strongly-interacting systems~\cite{cory1990time,mcdonald1989nmr}, not achieving optimal AC sensitivity, despite the pressing need for such protocols to further improve sensitivity in high density spin ensembles~\cite{acosta2009diamonds,mitchell2019sensor,mitchell2019quantum,otherpaper}. Here, we show that our framework addresses these challenges, by designing robust AC field-sensing pulse sequences that achieve maximal sensitivity to the target signal, while decoupling on-site disorder and spin-spin interactions. Furthermore, our formalism also provides a systematic approach to attain optimal sensitivity under given constraints, allowing diverse sensing strategies optimized for different scenarios.

\subsection{General Formalism for AC Magnetometry}
To achieve AC field sensing using interacting spin ensembles, we first incorporate external AC signals into the average Hamiltonian analysis. Specifically, the extra Hamiltonian due to the external AC signal can be modeled as
\begin{align}
H_\text{AC}(t)=\gamma B_\text{AC} \cos(2\pi f t - \phi) \sum_i S_i^z,
\end{align}
where $\gamma$ is the gyromagnetic ratio of the spins, $B_\text{AC}, f $ and $\phi$ are the amplitude, frequency, and phase of the target AC signal, respectively.

For a given pulse sequence represented as ${\bf F} = [F_{\mu,k}]$, we can apply the same average Hamiltonian analysis to understand how driven spins experience the sensing field in their effective frame, giving
\begin{align}
H_\text{avg,AC} =\gamma \vec{B}_\text{eff}(f) \cdot \sum_i \vec{S}_i \label{eq:avesense}
\end{align}
with
\begin{align}
	B_{\text{eff},\mu}(f) &= B_\text{AC} \Re \left[ e^{i\phi} \tilde{F}_\mu(f) \right], \label{eq:Beff}\\
	\tilde{F}_\mu(f) &= \frac{1}{T}\int_0^T e^{-i 2\pi f t}F_{\mu}(t)dt, \label{eq:Ftilde}
\end{align}
for $\mu = x,y,z$, and $\Re$ denotes the real part. Physically, the time-averaged sensing-field Hamiltonian [Eq.~(\ref{eq:avesense})] has a simple and elegant interpretation: in the toggling-frame picture, all driven spins will undergo a coherent precession around the effective magnetic field $\vec{B}_\text{eff}$. Additionally, as seen in Eq.~(\ref{eq:Beff}), the orientation and strength of $\vec{B}_\text{eff}$ are determined by the frequency-domain resonance characteristics of the applied pulse sequence, $\tilde{F}_\mu = \mathcal{F} [F_\mu]$, where $\mathcal{F}$ denotes the Fourier transform [Eq.~(\ref{eq:Ftilde})].

The AC magnetic field sensitivity $\eta_\text{ac}(f)$, characterizing the minimum detectable signal strength for an AC signal at frequency $f$, scales as
\begin{align}
	\eta_\text{ac}(f) \propto \frac{1}{\sqrt{T_2}|\tilde{F}_t (f)|}, 
\end{align}
where $|\tilde{F}_t (f)|$ is the total spectral response at the resonance frequency $f$ under the pulse sequence and $T_2$ is the coherence time of the spin ensemble. Physically, $|\tilde{F}_t (f)|$ can be understood as the effective signal strength experienced by the driven spins at resonance, namely $|\tilde{F}_t (f)| = |\vec{B}_\text{eff}(f)|/B_\text{AC}$, given by
\begin{align}
		|\tilde{F}_t(f) |&=\sqrt{\sum_\mu |\tilde{F}_\mu(f)|^2 \cos^2 (\phi- \tilde{\phi}_\mu(f))}. \label{eq:Beffmag}
\end{align}
Here, $\tilde{\phi}_\mu(f)$ is the spectral phase of $\tilde{F}_\mu(f)$ along the $\mu$-axis, identified from $\tilde{F}_\mu(f) = |\tilde{F}_\mu(f)| e^{-i \tilde{\phi}_\mu(f)}$. We immediately see that $|\tilde{F}_t(f) | \le \sqrt{\sum_\mu |\tilde{F}_\mu(f)|^2}$ from Eq.~(\ref{eq:Beffmag}), with the equality saturated when $\phi = \tilde{\phi}_x(f) =  \tilde{\phi}_y(f) =  \tilde{\phi}_z(f)$. Thus, it is crucial to align and synchronize the spectral phases of the pulse sequence at the target frequency $f$ to the phase of the sensing signal, in order to achieve the best sensitivity. In such a phase-synchronized case, the effective magnetic field becomes
\begin{align}
	\vec{B}_\text{eff}(f) = B_\text{AC} \left[|\tilde{F}_x(f)| \hat{e}_x + |\tilde{F}_y(f)| \hat{e}_y + |\tilde{F}_z(f)| \hat{e}_z \right].
\label{eq:Beffideal}
\end{align}
To optimally detect this effective sensing field, spins then need to be initialized perpendicular to $\vec{B}_\text{eff}$ to form the largest precession trajectory and maximize signal detection contrast. In addition, to optimize contrast for a projective measurement along the $\hat{z}$-axis, for readout the precession plane should be rotated to contain the $\hat{z}$-axis.

\subsection{Design Considerations for Efficient Quantum Sensing}
\label{sec:senseopt}
The additional requirements of optimizing magnetic field sensitivity impose new algebraic constraints within our framework. Here, we discuss the implications of these new constraints on the structure of sensing pulse sequences by utilizing the techniques described in Sec.~\ref{sec:Shortest}.

For efficient quantum sensing and to decouple on-site disorder as rapidly as possible, it is desirable to maintain a periodic structure in which the free evolution periods have frame directions that alternate between $+1, -1$, as adopted in the standard sensing sequence XY-8 as well as a new sequence Seq.~C we designed for interacting spin ensembles (see Fig.~\ref{fig:fig6}). However, for {\it interacting} ensembles where interaction-symmetrization is performed, this implies that any interface between two frame orientations will always have a fixed odd parity, and will thus violate condition 3 in Tab.~\ref{tab:summary} if single $\pi/2$ pulses are used for the frame transformations. Thus, to preserve sequence robustness, it is necessary to use {\it composite} pulse structures in which each frame-switching rotation is realized by a combination of two $\pi/2$ pulses to intentionally inject even parities to counteract the odd parities. An example of such a composite pulse is shown in Fig.~\ref{fig:fig6}(b).

The new sensing sequence, Seq.~C, has identical spectral responses between different axes, $|\tilde{F}_x(f)| = |\tilde{F}_y(f)| = |\tilde{F}_z(f)|$, leading to the transformation of a bare $\hat{z}$-axis resonant sensing field into the [1,1,1]-directional effective field, $\vec{B}_\text{eff}$, in the average Hamiltonian picture. While $\vec{B}_\text{eff} \parallel [1,1,1]$ is close to optimal for interacting ensembles, its strength can be further improved by adding an imbalance in the effective phase accumulation along each axis. Although the sum of the phase accumulation along all axes is fixed, the effective field strength depends on the sum of {\it squares} of the phase accumulation [Eq.~(\ref{eq:Beffmag})]. Thus, due to this nonlinearity, the effective field strength can be increased when the phase accumulation is different along the three axes, which is achieved by choosing the frame along one of the axes to be at the maxima of the sinusoidal sensing signal, resulting in enhanced phase accumulation (see Appx.~\ref{suppsec:optimalsense} and Seq.~I in Fig.~\ref{fig:fig9} for details).

Utilizing these ideas, we demonstrate in Ref.~\cite{zhou2019quantum} a solid-state AC magnetometer operating in a new regime by surpassing the sensitivity limit imposed by spin-spin interactions at high densities. In addition, our average Hamiltonian approach also helps to identify other undesired effects, such as spurious harmonics~\cite{loretz2015spurious,wang2019randomization}, which appear as additional spectral resonances in the total modulation function for finite pulse duration. This clearly demonstrates the utility of our formalism for the design of quantum sensing pulse sequences in the presence of interactions, disorder, and control imperfections.

\section{Application: Quantum Simulation with Tunable Disorder and Interactions}
\label{sec:simulation}
Our framework can also be readily adapted to \textit{engineer} various Hamiltonians in the context of quantum simulation. Here, the goal is to realize different types of interactions with tunable on-site disorder via periodic driving~\cite{hayes2014programmable,ajoy2013quantum,bookatz2014hamiltonian,choi2017dynamical,okeeffe2019hamiltonian,lee2016floquet,haas2019engineering}, such that one can explore a range of interesting phenomena in out-of-equilibrium quantum many-body dynamics, including dynamical phase transitions~\cite{lindner2011floquet,jiang2011majorana,heyl2018dynamical}, quantum chaos~\cite{dalessio2016from,garttner2017measuring} and thermalization dynamics~\cite{nandkishore2015many,abanin2018many,alvarez2015localization,wei2018exploring,wei2018emergent,ho2017critical,choi2019probing}. Moreover, the interplay of disorder, interactions and periodic driving can also lead to novel nonequilibrium phases of matter, such as the recently-discovered discrete time crystals~\cite{khemani2016phase,else2016floquet,von2016absolute,yao2017discrete,choi2017observation,zhang2017observation,sacha2018time,rovny2018observation,pal2018temporal}.

\begin{figure}
\begin{center}
\includegraphics[width=\columnwidth]{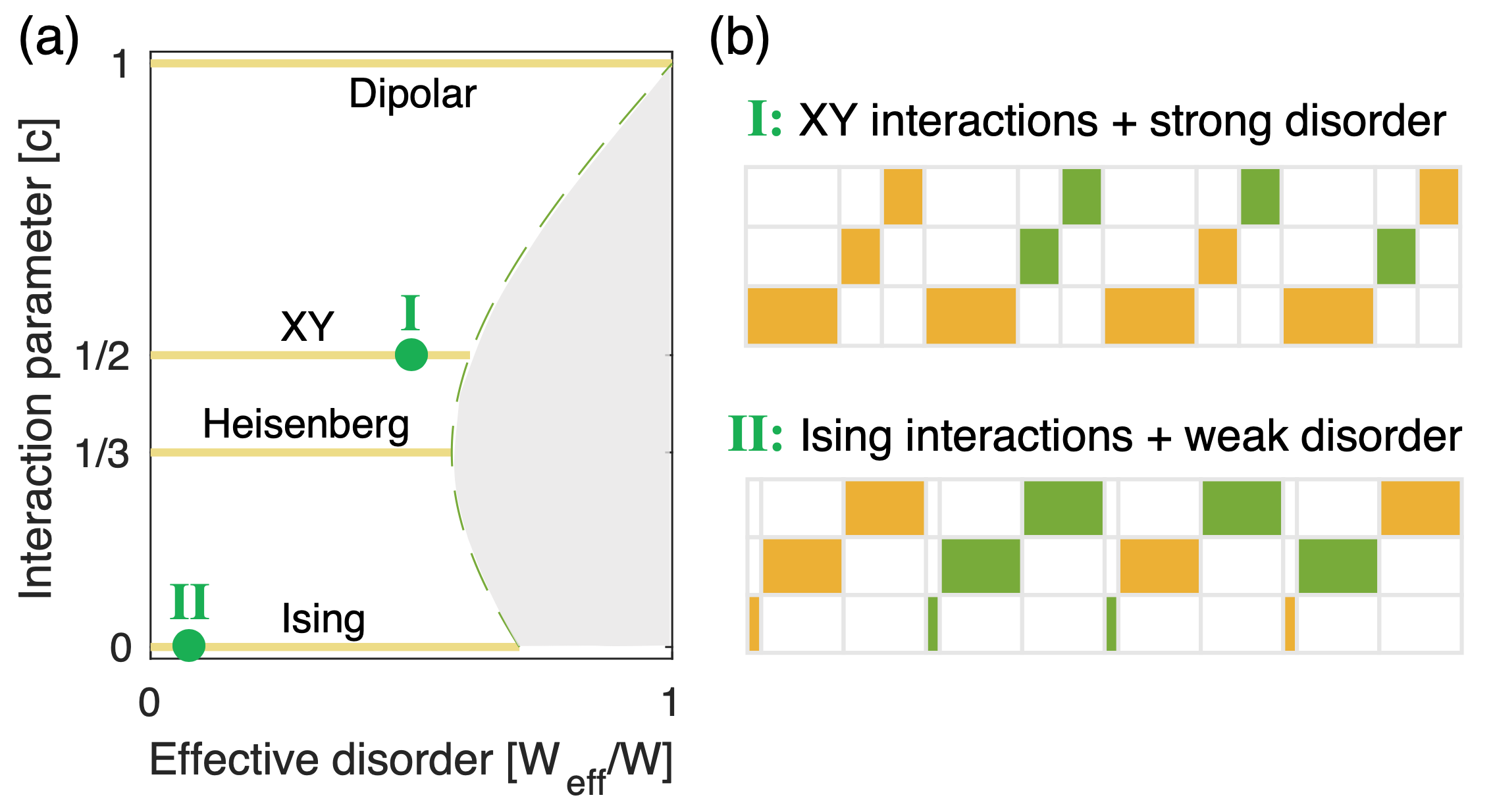}
\caption{{\bf Hamiltonian engineering for quantum simulations}. (a) Floquet-engineered many-body Hamiltonians with tunable disorder strengths and interaction types. The various interaction types including Ising ($c=0$), Heisenberg ($c=1/3$), XY ($c=1/2$) and dipolar interactions ($c=1$), can be realized by varying the single parameter $c$ which is defined as the relative fraction of the total time evolution along the $\hat{z}$-axis in the toggling frame. In this example, the system is chosen to be a disordered dipolar interacting spin ensemble with $J_{ij}^S = -J_{ij}^I$ and $J_{ij}^A = 0$, naturally realized with NV centers in diamond~\cite{kucsko2018critical}. The dashed line indicates the maximum effective disorder strength $W_\text{eff}$ achievable for a given interaction type, $W_\text{eff}/W = \sqrt{c^2 + (1-c)^2/2}$, where $W$ is the native on-site disorder strength. The gray area denotes the Hamiltonian regimes that are not accessible. (b) Examples of robust periodic pulse sequences that generate (I) XY interactions with strong disorder and (II) Ising interactions with weak disorder. These sequences correspond to the two green markers in (a).}
\label{fig:fig7}
\end{center}
\end{figure}

Indeed, as can be seen in Eqs.~(\ref{eq:Havedis}-\ref{eq:HaveAntiSym}), we can design the toggling-frame spin operators $\tilde{S}^z_k = \sum_{\mu} F_{\mu,k} S^\mu$ to achieve a nonzero target sum in Eqs.~(\ref{eq:rowsum},\ref{eq:absrowsum}) and engineer the leading-order average Hamiltonian $H_\text{avg}$.
More specifically, we show that for the common form of two-body interaction Hamiltonians $H_\text{int} = \sum_{ij} J_{ij}^S (S_i^x S_j^x + S_i^y S_j^y) + J_{ij}^I S_i^z S_j^z$, the relative strength between Ising and spin-exchange interactions can be tuned by the single parameter $c$ as
\begin{align}
	\tilde{J}_{ij}^S & = \frac{1+c}{2}J_{ij}^S + \frac{1-c}{2}J_{ij}^I, &\tilde{J}_{ij}^I & = (1-c) J_{ij}^S + c J_{ij}^I,
\label{eq:Jtuning}
\end{align}
where $c$ captures the imbalanced time evolutions in the toggling frames, defined such that the system evolves under the $\hat{z}$- and $\hat{x},\hat{y}$- axes toggling frames for total durations $cT$ and $(1-c)T/2$ in one sequence cycle $T$. Thus, the Floquet-engineered interaction Hamiltonian $H_\text{avg}^\text{int}$ now exhibits modified Ising and exchange interaction strengths of $\tilde{J}^I_{ij}$ and $\tilde{J}^S_{ij}$, respectively. Taking the case of interacting NV spin ensembles as an example~\cite{kucsko2018critical}, where $J_{ij}^S=-J_{ij}^I$, we can continuously interpolate between Ising ($c=0$), Heisenberg ($c=1/3$), XY ($c=1/2$), and dipolar-like ($c=1$) interactions by tuning the proportion $c$ of the sequence. Moreover, on-site disorder can also be independently controlled by introducing an additional sign imbalance along each axis in the toggling frame, changing the original disorder Hamiltonian $H_\text{dis} = \sum_i h_i S_i^z$ to the Floquet-engineered version $H_\text{avg}^\text{dis} = \sum_i (\vec{h}_\text{eff})_i \cdot \vec{S}_i$, where the effective disorder field $\vec{h}_\text{eff}$ can now have both longitudinal and transverse field components. 

We illustrate the accessible range of disorder and interaction Hamiltonians with this scheme in Fig.~\ref{fig:fig7}(a). Note that the maximum effective disorder strength $W_\text{eff}$ is dependent on $c$ (dashed line in Fig.~\ref{fig:fig7}(a), see Sec.~S1G~\cite{SM}). Two representative examples of how to engineer such interaction Hamiltonians in a robust fashion are shown in Fig.~\ref{fig:fig7}(b).

Combined with the techniques for robust engineering of other terms in the Hamiltonian, such that imperfections are suppressed, this allows access to a broad range of interacting, disordered Hamiltonians that potentially exhibit very different thermalization properties~\cite{nandkishore2015many,abanin2018many,alvarez2015localization,wei2018exploring,wei2018emergent,ho2017critical,choi2019probing}. Thus, our framework will open up a new avenue for the robust Floquet engineering of many-body Hamiltonians.

\section{Experimental Demonstration}
\label{sec:experiment}
\begin{figure}
\begin{center}
\includegraphics[width=\columnwidth]{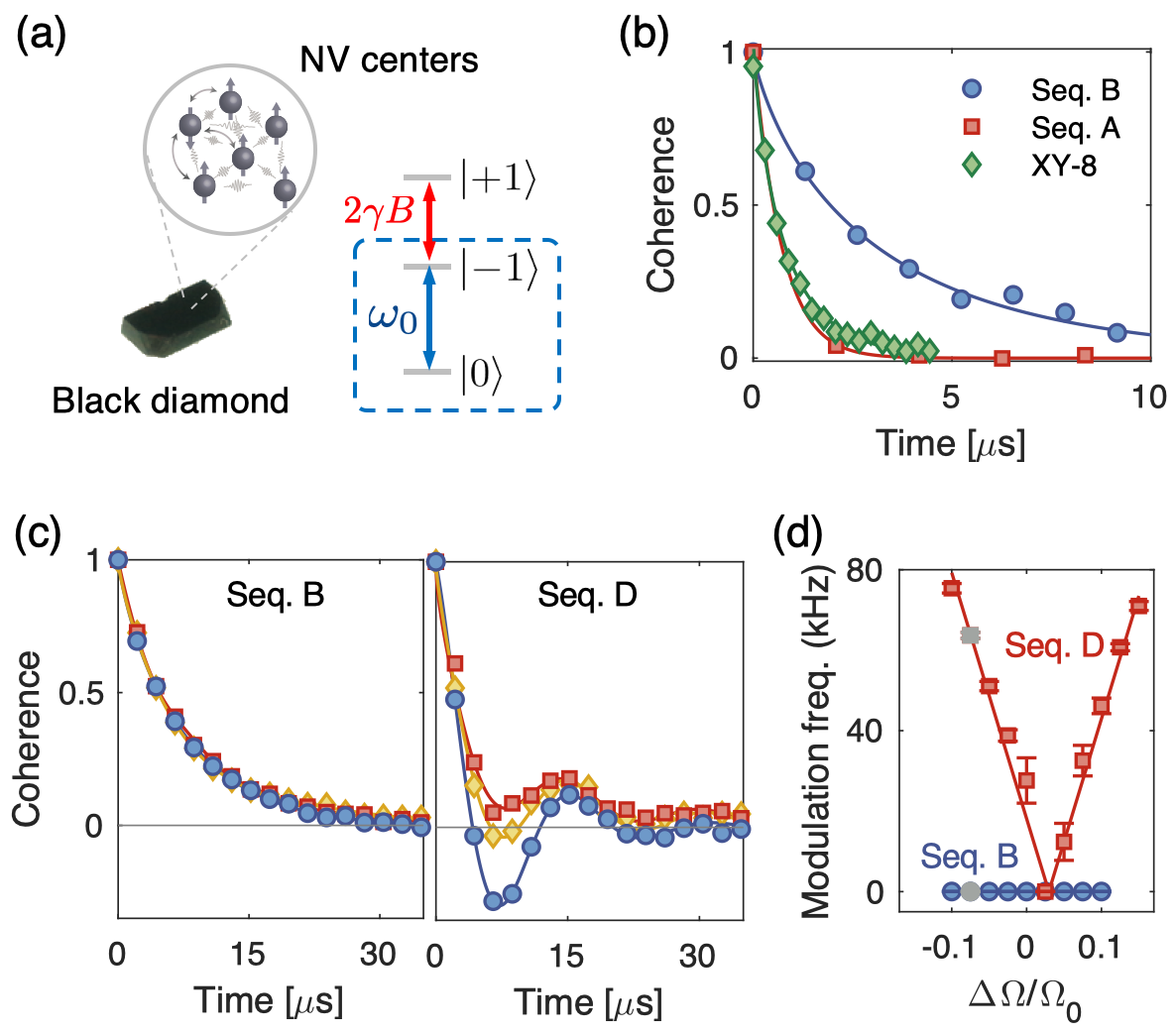}
\caption{{\bf Experimental demonstration.} (a) Sample used in experiments, containing strongly disordered, interacting NV centers in black diamond. We isolate two levels from the spin-1 ground state, $\{\ket{0},\ket{\pm1}\}$, by a static Zeeman shift (red arrow), to define an effective spin-1/2 (qubit) system, $\{\ket{0},\ket{-1}\}$. Resonant pulsed driving at frequency $\omega_0$ is applied to the spin to manipulate its quantum states. In the sample, NV centers at a distance interact with one another via magnetic dipolar interactions (circled diagram). (b) Comparison of coherence decay of driven spins under Seq.~A (Cory-48, red), Seq.~B (blue) and XY-8 (green). For sequence evaluation, we use two degenerate NV groups, which exhibit position-dependent random couplings with various interaction types including Ising, symmetric and anti-symmetric spin-exchange interactions~\cite{kucsko2018critical}, thus representing the most general class of one- and two-body interaction Hamiltonians. The pulse spacing and duration are fixed to $\tau = 15$~ns and $t_p =  6$~ns, respectively, and resulting decay profiles are fitted to a stretched exponential function. (c) Sequence robustness against spin-manipulation error. Here, we tested sequences on a single, isolated NV group. In the presence of systematic spin-rotation angle deviation $0.925(\pi/2)$ (gray points in (d)), the {\it non-robust} Seq.~D (right) shows modulations in the coherence profile for all three initial states polarized along the $\hat{x}$ (blue), $\hat{y}$ (red) and $\hat{z}$ (yellow) axes; the robust Seq.~B is insensitive to such errors, corroborated by the absence of the modulation. (d) Modulation frequency for Seq.~B and Seq.~D as a function of systematic rotation angle error relative to the perfect $\pi/2$ rotation, $\Omega_0 t_p = \pi/2$. The small lateral offset observed in Seq.~D at zero modulation frequency is due to a slight Rabi frequency calibration error.} 
\label{fig:fig8}
\end{center}
\end{figure}

Our framework is generally applicable to many different quantum systems, including interacting electronic spin ensembles, such as NV centers in diamond~\cite{doherty2013nitrogen,schirhagl2014nitrogen,awschalom2013quantum,dobrovitski2013quantum,koehl2011room}, phosphorus donors in silicon~\cite{tyryshkin2003electron,feher1959electron}, and rare earth ions~\cite{thiel2011rare}, conventional NMR systems~\cite{alvarez2015localization,wei2018exploring}, trapped ions~\cite{blatt2012quantum,jurcevic2014quasiparticle,bohnet2016quantum,zhang2017observation}, and even to emerging platforms of cold molecules~\cite{carr2009cold,yan2013observation,bohn2017cold} and Rydberg atom arrays~\cite{labuhn2016tunable,bernien2017probing}. These different systems will have a variety of competing energy scales and distinct interaction types that determine their dynamics, and will thus benefit from the flexibility of our system-targeted design formalism.

Here, we focus on the experimental implementation and demonstration of our results in an interacting ensemble of NV centers in diamond [Fig.~\ref{fig:fig8}(a)], tuned to realize the most general form of interactions, see Sec.~S2A~\cite{SM} and Refs.~\cite{choi2017depolarization,kucsko2018critical,otherpaper} for more details of our sample and experiments. It is characterized to be a disorder-dominated system [Fig.~\ref{fig:fig1}(a)], exhibiting large on-site disorder ($W \approx (2\pi) 4$ MHz) with modest interaction strengths ($J \sim (2\pi) 35$ kHz).

To demonstrate the wide applicability of our pulse sequence design formalism, we tune two groups of NV centers with different lattice orientations onto resonance with an external magnetic field. The corresponding Hamiltonian exhibits all different interaction types (Ising, symmetric and anti-symmetric spin exchange) with disordered, position-dependent coefficients, which represents the most general class of one- and two-body interaction Hamiltonians (see Ref.~\cite{kucsko2018critical} for more details). Despite the complex form of the interaction, system-targeted pulse sequences designed with our formalism enable a sizable extension of coherence times, as shown in Fig.~\ref{fig:fig8}(b). More specifically, we find that while a conventional XY-8 pulse sequence is limited by interactions to a coherence time of 0.9 $\mu$s, and Seq.~A (Cory-48) performs even worse in this parameter regime, our pulse sequence Seq.~B leads to an extension of the coherence time to 3.0 $\mu$s. This observation is consistent with the theoretical prediction [Fig.~\ref{fig:fig5}(c)] for a disorder-dominated system, and hence corroborates the importance of considering the energy hierarchy in designing dynamic decoupling pulse sequences.

In Fig.~\ref{fig:fig8}(c,d), to illustrate the importance of fulfilling the robust decoupling criteria [Tab.~\ref{tab:summary}], we compare the robustness of two different sequences to systematic rotation angle deviations. Here, we design a {\it non-robust} Seq.~D, which is almost identical to the robust Seq.~B, but does not suppress spin-rotation angle errors; In Seq.~D, intermediate frames are intentionally chosen to violate the suppression condition for rotation angle errors [Condition 4], while satisfying the rest of the conditions (see Fig.~\ref{fig:fig9} for more details of the sequence).

Fig.~\ref{fig:fig8}(c) illustrates the coherence decay profile of driven spins of a single, isolated NV group under these two sequences when the rotation angle is chosen to be 92.5$\%$ of the correct rotation angle (gray dots in Fig.~\ref{fig:fig8}(d)); Seq.~B does not show any oscillations, while Seq.~D shows pronounced oscillations over time, resulting from a residual error term $\delta H_\text{avg}$ (see Sec.~\ref{sec:angleerror}). This behavior is further confirmed in Fig.~\ref{fig:fig8}(d), where we extract the effective modulation frequency of the spin coherence as a function of the systematic rotation angle deviation. While Seq.~B does not show any oscillations, Seq.~D shows a linear dependence of oscillation frequency with the rotation angle error, indicating that it is not robust against perturbations.

\section{Discussion and Conclusion}
\label{sec:conclusion}
In this paper, we have introduced a novel framework for the efficient design and analysis of periodic pulse sequences to achieve dynamic Hamiltonian engineering that is robust against the main imperfections of the system. Our approach provides versatile means to design and adapt pulse sequences for a wide range of experimental platforms, by considering their system characteristics such as disorder, interactions and control inhomogeneities. Key to our approach is the adoption of a toggling frame description of the sequence and the resulting average Hamiltonian. Crucially, we find that various types of leading-order control errors can be systematically described by the time-domain transformations of a single interaction-picture Pauli spin operator during free evolution periods. This allows us to derive a simple set of algebraic conditions to fully describe all necessary conditions for specific target applications, significantly simplifying the design of pulse sequences. Remarkably, these algebraic conditions also allow the construction of efficient strategies and the proof of their optimality to enhance various figures of merit, such as sequence length and sensitivity. Furthermore, this approach can be readily interfaced with optimal control to substantially speed up the search of pulse sequences and take higher-order effects into account. Using a dense ensemble of interacting electronic spins in diamond, we experimentally confirm the wide applicability of our framework in systems with the most general form of one- and two-body interactions, thus confirming the generality of our approach.

In addition to its wide-reaching consequences on the systematic design and analysis of pulse sequences for various applications, our framework also opens up a number of intriguing directions for future studies. For example, we can extend our approach to higher-spin systems to investigate more complex quantum dynamics, such as quantum chaos and information scrambling, as well as utilize larger effective dipoles in those high-spin systems for more effective sensing~\cite{choi2017dynamical,okeeffe2019hamiltonian,fang2013high,mamin2014multipulse,bauch2018ultralong}. Higher-order contributions beyond the leading-order average Hamiltonian can also be systematically incorporated using the proposed framework. In addition, our formalism may also be extended to the synthesis of dynamically-corrected gates and other nontrivial quantum operations~\cite{khodjasteh2009dynamically,khodjasteh2009dynamical,khodjasteh2012automated}. While we have focused on the case of $\pi/2$ and $\pi$ pulses around $\hat{x}$, $\hat{y}$ axes for simplicity, it will be interesting to extend the analysis to more general control pulses, which could enable shorter protocols for Hamiltonian engineering. Moreover, by employing optimal control techniques to further boost the performance of the pulse sequences~\cite{khaneja2005optimal,doria2011optimal,iwamiya1993application,rose2018high}, we may be able to robustly engineer many-body Hamiltonians to create macroscopically entangled states, such as spin-squeezed states or Schr\"odinger-cat-like states, to be used as a resource for interaction-enhanced metrology beyond the standard quantum limit~\cite{cappellaro2009quantum,choi2018quantum}. 

\section*{Acknowledgements}
We thank P.~Cappellaro, W.~W.~Ho, C. Ramanathan, L.~Viola, F.~Machado for helpful discussions, and J.~Isoya, F.~Jelezko, S.~Onoda, H.~Sumiya for sample fabrication. We also thank A.~M.~Douglas for critical reading of the manuscript and assistance with numerical calculations. This work was supported in part by CUA, NSSEFF, ARO MURI, DARPA DRINQS, Moore Foundation GBMF-4306, Samsung Fellowship, Miller Institute for Basic Research in Science, NSF PHY-1506284.

\newpage
\appendix

\section{Average Hamiltonian Theory}
\label{sec:AHT}
Here, we introduce the basic principles of AHT and start by considering a generic time-dependent Hamiltonian for a driven quantum system
\begin{align}
H(t)=H_s+H_c(t),
\end{align}
where $H_s$ is the system Hamiltonian governing the internal dynamics and $H_c(t)$ describes the time-dependent control field used to coherently manipulate the spins (qubits). For a Floquet system, the control field is modulated in time with a periodicity of $T$, i.e., $H_c(t) = H_c(t+T)$.
At times $t=NT$, the many-body state is given by $\ket{\psi_{t=NT}} = \mathcal{U}(T)^N \ket{\psi_0}$ with the interaction picture unitary evolution operator~\cite{haeberlen1968coherent}
\begin{align}
	\mathcal{U}(T)=\mathcal{T}\exp\qty[-i\int_0^T \tilde{H}_s(t) \; dt],
\label{eq:exactH}
\end{align}
where $\mathcal{T}$ denotes time-ordering. Here, $\tilde{H}_s(t)$ is the rotated system Hamiltonian in the interaction picture with respect to control fields, given by $\tilde{H}_s(t) = U_c(t)^\dagger H_s U_c(t)$ with the unitary rotation operator $U_c(t) = \mathcal{T}\exp [-i\int_0^t H_c(t_1) \; dt_1 ]$. The control unitary rotation operator over one period is chosen to be identity $U_c(T)=\mathbb{I}$.

AHT allows the identification of a {\it time-independent} effective Hamiltonian $H_\text{eff}$ such that
\begin{align}
\mathcal{U}(T) = \exp [-i H_\text{eff} T].
\end{align}
The Magnus expansion of $\mathcal{U}(T)$ with expansion parameter $T$~\cite{magnus1954on} can be used to approximate this effective Hamiltonian as
\begin{align}
H_\text{eff} \approx \sum_{k=1}^{l} \bar{H}^{(k)},
\label{eq:Magnus}
\end{align}
where $l$ is the truncation order and $\bar{H}^{(k)}$ is the $k$-th order contribution in the Magnus expansion. The first two terms in the series are
\begin{align}
\bar{H}^{(0)}&=\frac{1}{T}\int_0^T \tilde{H}_s(t_1)dt_1\label{eq:AHT}\hspace{0.5 cm} \textrm{and}\\
\bar{H}^{(1)}&=-\frac{i}{2T}\int_0^T dt_2\int_0^{t_2}dt_1 \qty[\tilde{H}_s(t_2),\tilde{H}_s(t_1)].
\label{eq:Magnus1}
\end{align}
Although the accuracy of the average Hamiltonian approximation depends on the truncation order $l$, if the Floquet driving frequency $1/T$ is much faster than the local energy scales associated with the system Hamiltonian $H_s$, then the first few terms are sufficient to model $H_\text{eff}$ and approximate the dynamics of the many-body state to an accuracy improving exponentially in $l$~\cite{abanin2017effective,abanin2017rigorous,mori2016rigorous,kuwahara2016floquet}. In the following, we focus on the leading order contribution, corresponding to only retaining $\bar{H}^{(0)}$ in the series.

A general control field consists of $n$ pulses $\{P_{k=1,\cdots,n}\}$ with nonuniform pulse spacing $\{\tau_{k=1,\cdots,n}\}$, as shown in Fig.~\ref{fig:fig1}(d). Each $P_k$ defines a pulsed unitary rotation, generating a discrete set of rotated Hamiltonians $\{\tilde{H}_{k=1,\cdots,n}\}$, where 
\begin{align}
\tilde{H}_{k} = (P_{k-1} \cdots P_1)^\dagger H_s (P_{k-1} \cdots P_1)
\end{align}
with $\tilde{H}_1 = H_s$. As the interaction-picture Hamiltonian is rotated (toggled) at every pulse, $\tilde{H}_k$ are also referred to as the ``toggling-frame Hamiltonians" and govern the spin dynamics in their respective free evolution intervals $\tau_k$. For infinitely short pulses, the zeroth-order average Hamiltonian, $\bar{H}^{(0)} = H_\text{avg}$, can be simplified from an integral to a weighted average of the toggling-frame Hamiltonians, as presented in Eq.~(\ref{eq:Haverage}) of the main text.

\section{Details of Average Hamiltonian During Free Evolution Time}
\label{suppsec:aveideal}
In this section, we provide a detailed derivation of Eqs.~(\ref{eq:Havedis}-\ref{eq:HaveAntiSym}) characterizing the various average Hamiltonian contributions. The key idea is to express all Hamiltonian contributions in terms of rotationally-invariant terms and terms that only depend on the $S^z$ operator direction. Specifically, disorder and Ising interactions during the $k$-th free evolution period transform as
\begin{align}
S_i^z&\rightarrow \sum_\mu F_{\mu,k} S_i^\mu,\\
 S_i^z S_j^z&\rightarrow \sum_{\mu\nu}(F_{\mu,k} S_i^\mu)(F_{\nu,k} S_j^\nu)=\sum_\mu F_{\mu,k}^2 S_i^\mu S_j^\mu,
\end{align}
in the toggling-frame picture. Here we have used the fact that $F_{\mu,k}F_{\nu,k}=\delta_{\mu\nu} F_{\mu,k}^2$, since each column of the matrix $\mathbf{F} = [F_{\mu,k}]$ has only one nonzero element. Using these expressions, we can also easily find the transformed interaction for symmetric spin-exchange interactions by making use of the identity $S_i^xS_j^x+S_i^yS_j^y=\vec{S}\cdot\vec{S}-S_i^zS_j^z=(\sum_{\mu} S_i^\mu S_j^\mu)-S_i^zS_j^z$, which gives:
\begin{align}
S_i^xS_j^x+S_i^yS_j^y&\rightarrow \sum_{\mu} (1-F_{\mu,k}^2)S_i^\mu S_j^\mu.
\end{align}
Finally, we derive the transformation of the anti-symmetric spin-exchange interaction. To this end, we assume that the $S^x$ Pauli spin operator is transformed to $\tilde{S}^x(t)=\sum_\mu G_{\mu,k}S^\mu$, where $G_{\mu,k}$ satisfies the identity $G_{\mu,k}G_{\nu,k}=\delta_{\mu\nu}G_{\nu,k}^2$ and takes on values of $\{0,\pm 1\}$ (see Ref.~\cite{SM} for details of how one can explicitly construct $G_{\mu,k}$). Since the commutation relations between spin operators are conserved under frame transformations, the transformed $\tilde{S}^z(t)$ and $\tilde{S}^x(t)$ operators uniquely specify the $\tilde{S}^y(t)$ operator as:
\begin{align}
\tilde{S}^y(t)=\frac{1}{i}\qty[\tilde{S}^z(t),\tilde{S}^x(t)]=\sum_{\mu\nu\lambda}\epsilon_{\mu\nu\lambda}F_{\mu,k} G_{\nu,k} S^\lambda,
\label{eq:Sytilde}
\end{align}
where $\epsilon_{\mu\nu\lambda}$ is the Levi-Civita symbol. Based on this, we can write the transformation of the anti-symmetric spin-exchange interaction term as:
\begin{align}
S_i^xS_j^y-S_i^yS_j^x &\rightarrow \sum_{\mu\nu\lambda\sigma}G_{\sigma,k} \epsilon_{\mu\nu\lambda}F_{\mu,k} G_{\nu,k} (S_i^\sigma S_j^\lambda-S_i^\lambda S_j^\sigma)\nonumber\\
&=\sum_{\mu\nu\lambda}\epsilon_{\mu\nu\lambda}F_{\mu,k} G_{\nu,k}^2 (S_i^\nu S_j^\lambda-S_i^\lambda S_j^\nu)\nonumber\\
&=\sum_\mu F_{\mu,k}(\vec{S}_i\times \vec{S}_j)^\mu,
\end{align}
where we have used the identity presented above for $G_{\nu,k}$, as well as the fact that $G_{\nu,k}$ has only one nonzero element, squaring to 1. Combining these expressions gives the average Hamiltonian terms in Eqs.~(\ref{eq:Havedis}-\ref{eq:HaveAntiSym}).

\section{Analysis of Three-Body Interactions}
\label{suppsec:threebody}
In this appendix, we analyze the decoupling conditions for spin-1/2 three-body interactions in more detail. Interestingly, our versatile formalism can be applied to these more complex scenarios, leading us to a new set of rules that allow for the implementation of robust protocols in the presence of three-body interactions.

While most naturally occurring physical systems only involve two-body interactions, interactions involving more particles can lead to a number of exotic physical phenomena. For example, fractional quantum Hall state wavefunctions appear as the ground state of Hamiltonians involving three-body interactions~\cite{moore1991nonabelions,fradkin1998chern}, and many other topological phases and spin liquids are ground states of such many-spin Hamiltonians~\cite{moessner2001resonating,levin2005string}. There have also been various proposals for the direct realization of three-body interactions in experimental platforms ranging from cold molecules~\cite{buchler2007three} to superconducting qubits~\cite{mezzacapo2014many,chancellor2017circuit}. They may also emerge in the form of a higher-order term in the Magnus expansion of a system with only two-body interactions.

As a first step towards the control and engineering of such interactions, we analyze the conditions for dynamical decoupling for a polarized initial state. As in the main text, we will be focusing our attention on interactions under the secular approximation, where all terms in the Hamiltonian commute with a global magnetic field in the $\hat{z}$-direction.

\subsection{Ideal Pulse Limit}
We first prove a useful lemma for interactions under the secular approximation in the perfect, infinitely short pulse limit.

\textbf{Lemma:} For any interaction under the secular approximation, averaging over the spin-1/2 single qubit Clifford group is equivalent to averaging over toggling frames of $\tilde{S}^z$ that cover the six axis directions $\pm S^{x,y,z}$.

\textbf{Proof:} Consider a generic $N$-body interaction Hamiltonian $H$ and a set of unitary operators $U_k$ ($k=1,2,\cdots,K$). The average Hamiltonian over this set is given by
\begin{align}
H_\text{avg}=\frac{1}{K}\sum_{k=1}^K (U_k^\dagger)^{\otimes N}H U_k^{\otimes N}.
\end{align}
Let us now group the elements of the Clifford group into sets defined by how the elements transform the $S^z$ operator. Each set contains elements that satisfy $\tilde{S}^z=U^\dagger S^zU=(-1)^{\nu}S^\mu$ with $\nu = 0,1$, while the rotated $\hat{x}$-axis spin operator, $\tilde{S}^x = U^\dagger S^xU$, can take four distinct values that are orthogonal to the $\tilde{S}^z$ direction. In our toggling-frame representation, however, any of the four Clifford elements in the same set will correspond to a single term specified by $\tilde{S}^z$. Thus, proving the lemma reduces to proving that the four Clifford elements above give identical Hamiltonians.

We prove this by observing that for any two elements $U_1$ and $U_2$ in the same set, there exists a rotation $U_z$ around the $\hat{z}$ axis such that $U_1=U_zU_2$ (this rotation leaves the interaction picture $\tilde{S}^z$ invariant, but changes $\tilde{S}^x$). The Hamiltonian under conjugation by $U_1$ is then given by
\begin{align}
(U_1^\dagger)^{\otimes N}HU_1^{\otimes N}&=(U_2^\dagger)^{\otimes N}(U_z^\dagger)^{\otimes N}HU_z^{\otimes n}U_2^{\otimes N} \nonumber \\
&=(U_2^\dagger)^{\otimes N}HU_2^{\otimes N},
\end{align}
where we use $(U_z^\dagger)^{\otimes N}HU_z^{\otimes N} = H$. This holds because a rotation around the $\hat{z}$-axis does not modify the secular Hamiltonian, which commutes with the global $S^z$ operator. Consequently, a conjugation of the average Hamiltonian above by $U_1$ will be equal to a conjugation by $U_2$, and thus each set of Clifford elements that transform the $S^z$ operator in the same way will result in identical Hamiltonians. $\blacksquare$

With this lemma in hand, we can utilize mathematical results from unitary $t$-designs~\cite{divincenzo2002quantum,dur2005standard,emerson2003pseudo,collins2006integration,gross2007evenly,ambainis2007quantum,dankert2009exact,webb2015clifford,zhu2017multiqubit} to show that a polarized initial state will be an eigenstate of the three-body interacting Hamiltonian after symmetrization along the six axis directions (i.e. the $\pm \hat{x}, \pm \hat{y}, \pm \hat{z}$-axes), as described in Sec.~\ref{sec:multibody} of the main text. This is a consequence of the fact that the Clifford group is a unitary 3-design. However, as the Clifford group is not a unitary 4-design, four-body interactions will still induce dynamics after symmetrization. Indeed, we can explicitly verify this by considering the symmetrized interaction $(S^x)^{\otimes 4}+(S^y)^{\otimes 4}+(S^z)^{\otimes 4}$, which is found to act nontrivially on a generic polarized initial state.

\subsection{Finite Pulse Duration Effects}
We now illustrate how to analyze finite pulse duration effects for three-body interactions using our sequence representation matrix {\bf F}. We consider generic interaction Hamiltonians with up to three-body interactions and, in analogy to Eqs.~(\ref{eq:Havedis}-\ref{eq:HaveAntiSym}), we write the $k$-th toggling-frame Hamiltonian as a polynomial in $F_{\mu,k}$:
\begin{align}
\tilde{H}_k=\sum_\mu\sum_{l=0}^3 F_{\mu,k}^l \mathcal{O}_{\mu,l},
\end{align}
where $\mathcal{O}_{\mu,l}$ describes the generic operator form of interactions that transform as the $l$-th power of $F_{\mu,k}$.
More specifically, $\mathcal{O}_{\mu,l}$ can be written as a sum of terms, each composed of a product of Pauli operators that preserve the total magnetization and thus include an even number of $S^x$ or $S^y$ operators. Consequently, the interaction must either be of the form $S^zS^zS^z$, or involve the tensor product of an $S^z$ operator and a polarization-conserving two-body operator, which can be $S^xS^x+S^yS^y$ or $S^xS^y-S^yS^x$. Each of these terms can thus be written as a product of individual components that transform as $F_{\mu, k}$. For example, we can rewrite the following three-body interaction as
\begin{align}
&(S^xS^y-S^yS^x)S^z\nonumber\\=&\sum_{\mu\nu\sigma}\qty[\epsilon_{\mu\nu\sigma}F_{\mu, k} \qty(S^\nu S^\sigma-S^\sigma S^\nu)]\qty[F_{\mu, k}S^\mu].
\label{eq:l2}
\end{align}
This is a three-body interaction with $l=2$, since it is proportional to the square of $F_{\mu,k}$. This suggests that during the finite pulse duration between free evolution blocks $k$ and $k+1$, where the interaction-picture operator $\tilde{S}^z(\theta)=\sum_\mu(\cos\theta F_{\mu,k}+\sin\theta F_{\mu,k+1})S^\mu$ with $\theta$ evolving from 0 to $\pi/2$, the corresponding average Hamiltonian can be written as
\begin{align}
\bar{H}_{P_k}^{(0)}=\sum_{l}\frac{2}{\pi}\int_0^{\pi/2}d\theta \left[\sum_\mu\left(\cos\theta F_{\mu,k}+\sin\theta F_{\mu,k+1}\right)\right]^l \tilde{\mathcal{O}},
\label{eq:poly}
\end{align}
where $\tilde{\mathcal{O}}$ contains operators acting during the rotation pulse and is implicitly dependent on $l$ and the indices in the bracket (for example, when $\theta=0$, $\tilde{\mathcal{O}} = \mathcal{O}_{\mu,l}$).

Expanding the polynomial in the bracket of Eq.~(\ref{eq:poly}), we shall find contributions corresponding to terms of degree $u$ in $F_{\mu,k}$ and $v$ in $F_{\mu,k+1}$, with $u+v=l$, since $F_{\mu,k}$ and $F_{\mu,k+1}$ each have only one nonzero element. This allows us to generalize the conditions for decoupling finite pulse imperfections discussed in the main text to three-body interactions, and also provides an alternative perspective to the conditions in the main text.
For interaction terms exhibiting linear ($l=1$) or quadratic ($l=2$) dependence on $F_{\mu,k}$, the decoupling conditions presented in Tab.~\ref{tab:summary} of the main text can be directly applied.
Similarly, for $l=3$, the terms $F_{\mu,k}^3$ and $F_{\mu,k+1}^3$ directly correspond to the three-body interactions appearing in the original Hamiltonian within the free evolution periods $k$ and $k+1$, and can be easily incorporated into the sequence design by extending the effective duration of the free evolution time.
Meanwhile, the decoupling of cross terms $F_{\mu,k}^2F_{\nu,k+1}$ and $F_{\mu,k}F_{\nu,k+1}^2$ correspond to a generalization of the interaction cross-term decoupling condition [Condition 3 in Tab.~\ref{tab:summary}] described in the main text:
\begin{align}
 \sum_k F_{\mu,k}^2F_{\nu,k+1}+F_{\mu,k+1}^2F_{\nu,k}=0,  
\end{align}
for each pair of $(\mu,\nu)$. As an example, for the pair of directions $\hat{x}$, $\hat{z}$, we consider all instances in which the $\hat{x}$ and $\hat{z}$ frames appear in the free evolution frames immediately preceding and following a $\pi/2$ pulse, and the above decoupling condition requires that the signs of all $\hat{x}$ frames appearing in such positions sum up to 0.

Combining these results, we see that our formalism provides a systematic method to robustly decouple the effects of any three-body interaction under the secular approximation, on any polarized initial state, even in the presence of finite pulse durations.

\begin{figure*}
\begin{center}
\includegraphics[width=2\columnwidth]{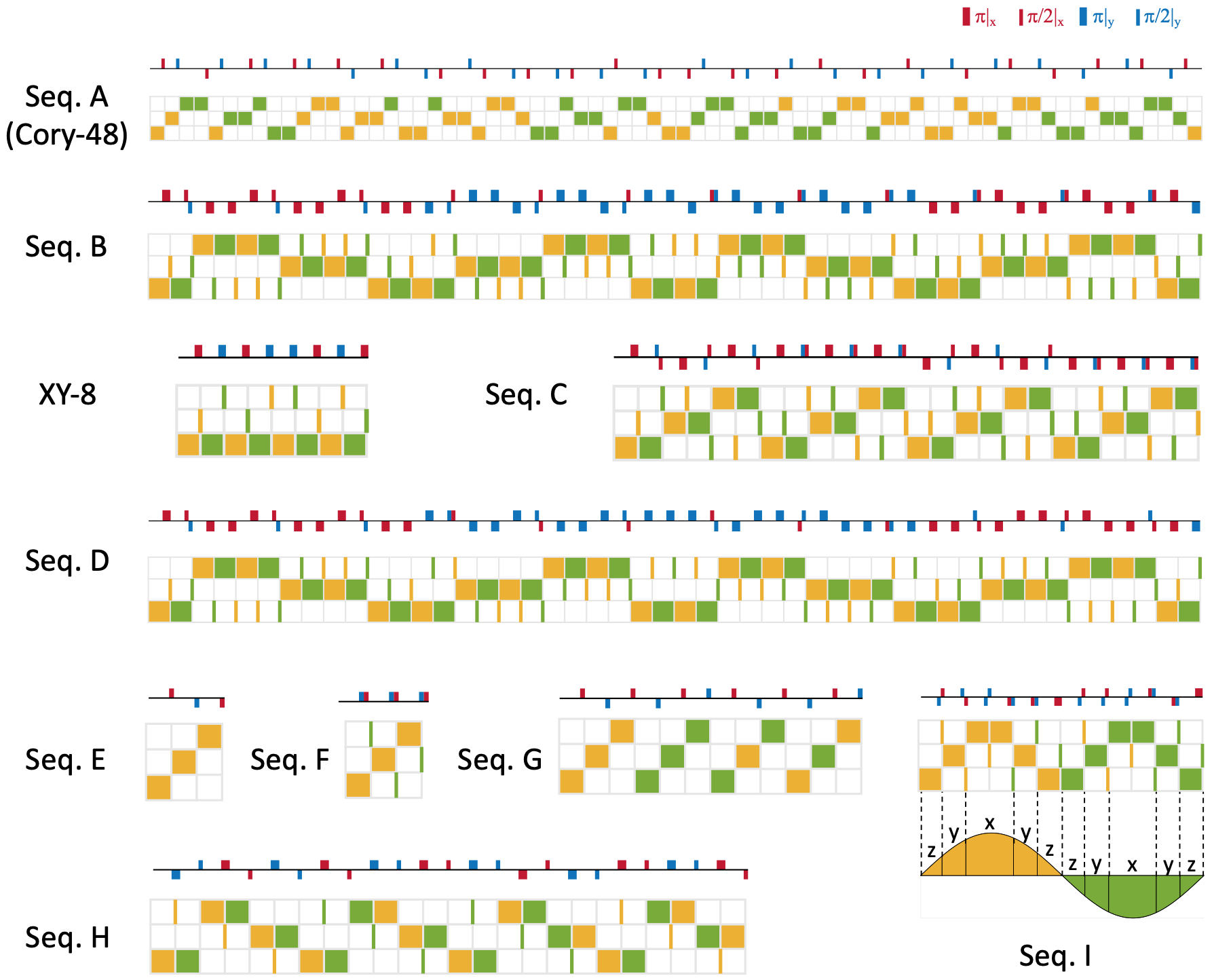}
\caption{{\bf Complete representation of periodic pulse sequences.} All sequences are shown in the conventional pulse representation and our toggling-frame transformation-based representation. In the toggling frame representation, squares indicate free evolution times of length $\tau$ and narrow lines indicate short intermediate frames; yellow is positive and green is negative. Seq.~A: The Cory-48 sequence, which decouples interactions on a faster timescale and disorder on a slower timescale, and is robust against leading-order imperfections; Seq.~B: Pulse sequence designed to decouple disorder on a faster timescale and interactions on a slower timescale, and is robust against leading-order imperfections; XY-8: Standard pulse sequence for dynamical decoupling with non-interacting spins; Seq.~C: Pulse sequence to illustrate robust dynamical decoupling of interactions and disorder, and well-aligned sensing resonances; Seq.~D: Pulse sequence that has identical free evolution frames as Seq.~B, but with intermediate frames in the second half permuted to remove the robustness against rotation angle errors. Seq.~E: Simple sequence to fully symmetrize interactions; however, the transverse spin operators do not return to themselves at the end of this sequence; Seq.~F: Modified version of Seq.~E, in which the composite pulses ensure that the transverse spin operators return at the end of the sequence; Seq.~G: Minimal pulse sequence that meets all dynamical decoupling and robustness requirements (Tab.~\ref{tab:summary} in the main text) without composite pulses; Seq.~H: Minimal pulse sequence that meets all dynamical decoupling and robustness requirements without composite pulses, and has a fast spin echo structure; Seq.~I: Minimal pulse sequence that achieves optimal vector sensitivity under interaction-decoupling constraints, and illustration of phase accumulation along each axis.}
\label{fig:fig9}
\end{center}
\end{figure*}

\section{Efficient Sequence Design Strategies}
\subsection{Minimal Length for Robust Dynamical Decoupling}
\label{suppsec:optimal}
In this section, we discuss the minimal sequence lengths required to satisfy different combinations of decoupling conditions, and provide examples of pulse sequences that achieve these minimal lengths.

We start by considering the minimal number of free evolution blocks to fully symmetrize the interaction Hamiltonian. From condition 2 of Tab.~\ref{tab:summary}, we see that the minimal nontrivial solution requires nonzero elements in each of the three rows of ${\bf F}$. Consequently, at least 3 free evolution blocks are required. A possible realization is (see also Seq.~E in Fig.~\ref{fig:fig9})
\begin{align}
\begin{pmatrix}
{\bf F}\\
\boldsymbol{\tau}
\end{pmatrix}=
\begin{pmatrix}
0 & 0 & 1\\
0 & 1 & 0\\
1 & 0 & 0\\
\tau & \tau & \tau
\end{pmatrix}.
\end{align}
However, according to the discussion in Sec.~S1E~\cite{SM}, a single cycle evolution under this sequence does not return the transverse spin operators to their original configuration, i.e. $U_c(t)\neq \mathbb{I}$. To achieve $U_c(t) = \mathbb{I}$, we can use
\begin{align}
\begin{pmatrix}
\mathbf{F}\\\boldsymbol{\tau}
\end{pmatrix}
=\begin{pmatrix}
0 & -1 & 0 & 0 & 1 & 0\\
0 & 0 & 1 & 0 & 0 & -1\\
1 & 0 & 0 & -1 & 0 & 0\\
\tau & 0 & \tau & 0 & \tau & 0\\
\end{pmatrix},
\end{align}
with composite $\pi/2$ pulses inserted at the interfaces of free evolution intervals (see Seq.~F in Fig.~\ref{fig:fig9}). However, neither of the above sequences are robust against on-site disorder or finite pulse duration effects.

As discussed in Sec.~\ref{sec:Shortest} in the main text, to fully symmetrize interactions and cancel disorder, at least 6 free evolution periods are required. If composite pulses are allowed, there is a pulse sequence consisting of 6 free evolution periods that also satisfies all robustness requirements, as described in Sec.~\ref{sec:Shortest} the main text. In some scenarios, however, composite pulses may be undesirable due to the technical challenges of implementing independent pulses in quick succession; in this case, we can show that at least 12 free evolution intervals are required. Let us first examine why 6 intervals are not sufficient: The cross-interaction parity condition [Condition~3 in Tab.~\ref{tab:summary}] requires the number of odd-parity frame changes to be equal to that of even-parity frame changes; a sequence with 6 free evolution intervals thus gives 3 even-parity and 3 odd-parity frame changes. The odd number of odd-parity frame changes, however, results in a toggling-frame operator at the beginning of the next cycle $\tilde{S}^z(T)$ that has opposite sign from $\tilde{S}^z(0)$, violating the periodic condition $\tilde{S}^z(0) = \tilde{S}^z(T)$. To prevent this, we thus require an {\it even} number of odd-parity frame changes. Since the length of the sequence has to be an integer multiple of 6 to simultaneously accomplish disorder and interaction decoupling, the minimal sequence length is at least 12 frames, which now realizes a valid Floquet cycle with $U_c(T)=\mathbb{I}$ and satisfies all decoupling conditions. 

One realization of such a pulse sequence, for example, can be written in our representation as (see also Seq.~G in Fig.~\ref{fig:fig9})
\begin{align}
\begin{pmatrix}
{\bf F} \\
\boldsymbol{\tau}
\end{pmatrix}_\text{Opt-12}
=\left(\begin{array}{ccccccccccccc}
0 & 0 & 1 & 0 & 0 & -1 & 0 & 0 & -1 & 0 & 0 & 1 \\
0 & 1 & 0 & 0 & -1 & 0 & 0 & 1 & 0 & 0 & -1 & 0 \\
1 & 0 & 0 & -1 & 0 & 0 & -1 & 0 & 0 & 1 & 0 & 0\\
\tau & \tau & \tau & \tau & \tau & \tau &\tau & \tau & \tau & \tau & \tau & \tau\\
\end{array}\right) \nonumber.
\end{align}
The above argument can also be readily extended to other scenarios. For disorder-dominated systems it is desirable to maintain a fast spin-echo structure on the toggling-frame evolution of a sequence, as discussed in Sec.~\ref{sec:decouple}. This necessitates a frame matrix structure in which the frames along each axis always consist of pairs with opposite sign. When permitting only a single $\pi/2$ pulse to switch frames, the parity product of the last element of the first pair and the first element of the second pair will again have the same parity constraints as the preceding case. Thus, we can apply the same argument to show that the 12-frame sequence also needs to be doubled to satisfy the parity condition if composite pulses are not allowed and a spin-echo structure is required on a fast timescale. One example that achieves this is illustrated as Seq.~H in Fig.~\ref{fig:fig9}.

\subsection{Composite Pulses for Sensing}
\label{suppsec:optimalsense}
We now discuss the implications of the algebraic conditions in Tab.~\ref{tab:summary} in the context of quantum sensing. Here, we show that for pulse sequences that follow a periodic sign-modulation structure along each axis (for instance, the sequences in Fig.~\ref{fig:fig6} of the main text), to fully utilize the effective sensing field and satisfy the robust dynamical decoupling conditions, it is necessary to employ composite pulses for the effective $\pi/2$-pulse implementations to suppress interaction cross-terms [Condition 3].

If only single $\pi/2$ and $\pi$ pulses are employed to connect different axes in the toggling frame, the fixed spin-echo-type sign modulation patterns inevitably result in fixed parities at every interface between two axis directions, leading to the violation of condition 3 in Tab.~\ref{tab:summary} (see also discussion in Appx.~\ref{suppsec:optimal}). To address this, composite pulses should be utilized instead to connect different axes and adjust the parities, allowing one to balance the number of even and odd parity interfaces.

\newpage
\bibliography{main}

\end{document}


\title{Supplementary Materials for ``Robust Dynamic Hamiltonian Engineering of Many-Body Spin Systems"}
\date{\today}

\affiliation{Department of Physics, Harvard University, Cambridge, Massachusetts 02138, USA}
\affiliation{School of Engineering and Applied Sciences, Harvard University, Cambridge, Massachusetts 02138, USA}
\affiliation{Department of Physics, University of California Berkeley, Berkeley, California 94720, USA}

\author{Joonhee Choi$^{1,2}$}
\thanks{These authors contributed equally to this work} 
\author{Hengyun Zhou$^{1}$}
\thanks{These authors contributed equally to this work} 
\author{Helena S. Knowles$^{1}$}
\author{Renate Landig$^{1}$}
\author{Soonwon Choi$^{3}$} 
\author{Mikhail D. Lukin$^{1}$}
\email{lukin@physics.harvard.edu}
\maketitle

\tableofcontents

\section{Formalism details}
\label{suppsec:hamiltonian}

\subsection{General Hamiltonian Under the Secular Approximation}
\label{suppsec:secular}
In this section, we discuss the consequences of the secular approximation (i.e. rotating wave approximation under a strong quantizing field), and explain why under this approximation the transformation properties of the $S^z$ spin operator uniquely specify the Hamiltonian.

The secular approximation is widely used in a large variety of contexts. Since the strong quantizing field along the $\hat{z}$ axis sets a dominant energy scale in the system Hamiltonian, under the secular approximation we only retain Hamiltonian contributions that conserve the total spin polarization along the $\hat{z}$ axis; in other words, a change in the total $\hat{z}$-axis polarization, such as a single spin flip, is associated with a large energy cost and is thus energetically suppressed. In mathematical terms, this approximation corresponds to only keeping Hamiltonian contributions that commute with $S^z_\text{tot} = \sum_i S^z_i$. This allows us to easily write down all allowed terms in the Hamiltonian for one- and two-body operators in the Pauli operator basis. For one-body operators, the only nontrivial term is $S^z$, since the other Pauli operators, $S^{x,y}$, do not commute with the $S^z_\text{tot}$ operator. Similarly, the only two-body operators commuting with $S^z_\text{tot}$ are the following three types: 
\begin{align}
S_i^z S_j^z, \quad \frac{1}{2}(S_i^+S_j^-+S_i^-S_j^+), \quad \frac{i}{2}(S_i^+S_j^--S_i^-S_j^+),
\end{align}
where $S^{\pm} = S^x \pm i S^y$. Hence, a linear combination of the one- and two-body operators allowed in the secular approximation results in the following general form of many-body Hamiltonians
\begin{align}
H_s&=\sum_i h_i S_i^z+ \sum_{ij}\left[J^{I}_{ij} S_i^zS_j^z + \frac{J^S_{ij}}{2}(S_i^+S_j^-+S_i^-S_j^+) +\frac{iJ^A_{ij}}{2}(S_i^+S_j^--S_i^-S_j^+)\right],
\end{align} 
which corresponds to Eq.~(2) of the main text.

More importantly, when global spin manipulation is performed, the $S^z_\text{tot}$-preserving nature of the interaction Hamiltonian under the secular approximation also greatly simplifies the representation of the Hamiltonian. To describe how a fully generic Hamiltonian is transformed after global rotations, we usually would need to specify how both $S^z$ and $S^x$ are transformed. However, due to the fact that the Hamiltonian commutes with the global $S^z_\text{tot}$ operator, one can perform a rotation with $S^z$ to take the $S^x$ operator into any direction that is perpendicular to the transformed $S^z$ direction, and the Hamiltonian remains invariant. Consequently, the Hamiltonian is uniquely specified given the transformation properties of the $S^z$ operator alone. This justifies our approach in the main text, where, given a Hamiltonian under the secular approximation, we only need to examine the transformation properties of the $S^z$ operator. 

While we have focused on the case of Hamiltonians satisfying the secular approximation, it would also be interesting to extend the techniques we have developed in this paper to situations where such an approximation does not hold, such as zero-field NMR~\cite{theis2011parahydrogen}.

\subsection{Details of Finite Pulse Width Analysis}
\label{suppsec:avefinite}
Here, we provide more details for the derivation of the average Hamiltonian for a $\pi/2$ pulse with finite pulse width. For illustration purposes, we consider a concrete example in which the initial toggling frame with $\tilde{S}^x(0)=S^x$, $\tilde{S}^y(0)=S^y$, $\tilde{S}^z(0)=S^z$ is rotated, such that the rotation brings the frame from $+\hat{z}$ into $+\hat{y}$. Correspondingly, in the interaction picture relative to the instantaneous control pulse, the operators are transformed as $\tilde{S}^x(\theta)=S^x$, $\tilde{S}^y(\theta)=S^y\cos\theta-S^z\sin\theta$, $\tilde{S}^z(\theta)=S^z\cos\theta+S^y\sin\theta$, where $\theta$ smoothly changes from 0 to $\pi/2$ over the finite duration of the pulse. Plugging this into the form of the average Hamiltonian, we find that the instantaneous toggling-frame Hamiltonian during the rotation is
\begin{widetext}
\begin{align}
\tilde{H}(\theta)
&=\sum_i h_i[S_{i}^z\cos\theta+S_{i}^y\sin\theta]+\sum_{i,j} J_{ij}^A[(S_{i}^xS_{j}^y-S_{i}^yS_{j}^x)\cos\theta+(-S_{i}^xS_{j}^z+S_{i}^zS_{j}^x)\sin\theta]\nonumber\\
&+\sum_{i,j} J_{ij}^S\vec{S}_i\cdot\vec{S}_j+\sum_{i,j} (J_{ij}^I-J_{ij}^S)[S_i^zS_j^z\cos^2\theta+S_i^yS_j^y\sin^2\theta + (S_i^yS_j^z+S_i^zS_j^y)\sin\theta\cos\theta].
\end{align}
\end{widetext}
Upon careful inspection, we see that the transformation properties of the Hamiltonian terms during the free evolution period (see Sec.~II in the main text) also dictate the response during the finite pulse width evolution. More specifically, the disorder and imaginary spin exchange terms, which transform linearly in $F_{\mu, k}$, are proportional to $\sum_{\mu,\nu}(\cos\theta F_{\mu, k}+\sin\theta F_{\nu,k+1})$ during the finite pulse duration. Subsequently, after integrating over the angle $\theta$, we arrive at Hamiltonian contributions that are proportional to $F_{\mu, k}$ and $F_{\nu, k+1}$, with proportionality factor $2/\pi$ due to the integration. Meanwhile, the symmetric spin exchange and Ising terms, which transform quadratically in $F_{\mu,k}$, will produce contributions proportional to $[\sum_{\mu,\nu}(\cos\theta F_{\mu, k}+\sin\theta F_{\nu,k+1})]^2$ during the finite pulse duration. After integrating over the angle $\theta$, this will result in Hamiltonian contributions that are proportional to $F_{\mu,k}^2$ and $F_{\nu,k+1}^2$ with factor $1/2$, and a cross-term proportional to $F_{\mu,k}F_{\nu,k+1}$ with factor $1/\pi$. Combining these considerations gives rise to the expression in Eq.~(20) of the main text. While we have assumed that the Rabi frequency remains fixed between different pulses, our framework can be easily generalized to incorporate differing Rabi frequencies, simply by specifying a different pulse width for each pulsed rotation.

The above derivation also provides a simple intuition for the different components of the finite pulse duration effect Hamiltonian: there will be terms that are proportional to the average Hamiltonian in the free evolution periods preceding and following the pulse, but for terms that transform quadratically, the square during the finite pulse duration will also generate cross terms. The same intuition will also prove to be useful when deriving conditions for finite pulse duration effects with three-body interactions (see section IV.3 of the main text).

\subsection{Analysis of Rotation Angle Errors}
\label{suppsec:angleerror}
We now consider the effects of control errors, in the form of a systematic rotation angle deviation. We will show that the effective Hamiltonian corresponding to the rotation angle error is well described by the chirality of frame changes, thus enabling a simple algebraic description of this decoupling condition.

To be more concrete, consider a $\pi/2$-pulse $P_k=\prod_i \exp\qty(-i\frac{\pi}{2}\sum_\mu \alpha_{\mu,k}S_i^\mu)$, where $\alpha_{\mu,k}$ specifies the rotation axis of the $k$-th pulse (e.g. $\alpha_{x,k}=+1$ implies a $+\hat{x}$-rotation). A rotation angle error corresponds to an actual rotation $P_k+\delta P_k=\prod_i \exp\qty(-i (\frac{\pi}{2}+\epsilon_i)\sum_\mu \alpha_{\mu,k}S_i^\mu)$ being applied, where $\epsilon_i$ is the angle error for the spin at site $i$, assumed to be static and independent of the applied pulse $P_k$; equivalently, we can regard this as an error term in the engineered Hamiltonian
\begin{align}
\delta H_{P_k}=\frac{1}{t_p} \sum_{i} \epsilon_i \sum_\mu \alpha_{\mu,k}S_i^\mu
\end{align}
acting during the rotation. The toggling-frame Hamiltonian corresponding to this rotation angle error can be calculated as 
\begin{align}
\delta \tilde{H}_{P_k}&=U_{k-1}^\dagger \delta H_{P_k} U_{k-1} \nonumber \\
&=\frac{1}{t_p} \sum_i \epsilon_i U_{k-1}^\dagger \qty(\sum_\mu \alpha_{\mu,k}S^\mu_i)U_{k-1},
\label{eq:Pkerr}
\end{align}
where $U_{k-1}=P_{k-1}\cdots P_2P_1$ is the unitary rotation due to the preceding control pulses. We simplify Eq.~(\ref{eq:Pkerr}) by making use of the expressions for the initial and final frames connected via the pulse $P_k$:
\begin{align}
U_{k-1}^\dagger S_i^z U_{k-1}&=\sum_\mu F_{\mu,k} S_i^\mu,\label{eq:Fk-1}\\
U_{k-1}^\dagger P_k^\dagger S_i^z P_k U_{k-1}=U_k^\dagger S_i^zU_k&=\sum_\mu F_{\mu,{k+1}}S_i^\mu.\label{eq:Fk}
\end{align}
Inserting $U_{k-1}U_{k-1}^\dagger=I$ into the latter equation, we arrive at the expression
\begin{align}
\qty(U_{k-1}^\dagger P_k U_{k-1})^\dagger\qty(\sum_\nu F_{\nu,k}S_i^\nu)\qty(U_{k-1}^\dagger P_k U_{k-1})
=\sum_\mu F_{\mu,k+1}S_i^\mu.
\end{align}
While this equation looks complicated, upon closer inspection, it implies that $U_{k-1}^\dagger P_kU_{k-1}$ is nothing other than the rotation pulse $\tilde{P}_k = \prod_i \exp\qty(-i\frac{\pi}{2}\sum_\mu \beta_{\mu,k}S_i^\mu)$ that in the toggling-frame representation, rotates the $k$-th toggling frame $\vec{F}_k$ into the $(k+1)$-th one $\vec{F}_{k+1}$, where $\vec{F}_k =\sum_\mu F_{\mu,k} \hat{e}_\mu$. The rotation axis direction in the toggling frame specified by $\beta_{\mu,k}$ can be simply obtained from a cross product between the toggling-frame orientations before and after the rotation:
\begin{align}
\vec{\beta}_k=\sum_\mu \beta_{\mu,k}\vec{e}_\mu=\vec{F}_{k+1}\times\vec{F}_{k},
\label{eq:defbeta}
\end{align}
where $\vec{\beta}_k$ characterizes the chirality of the frame change between neighboring $k$-th and $k+1$-th toggling frames.

Since $P_k=\prod_i \exp\qty(-i\frac{\pi}{2}\sum_\mu \alpha_{\mu,k}S_i^\mu)$, where $\sum_\mu \alpha_{\mu,k}S^\mu$ is a rotation along the $\hat{x}$ or $\hat{y}$ axis and is thus proportional to a Pauli matrix, we can move the unitary $U_{k-1}$ into the exponential
\begin{align}
U_{k-1}^\dagger P_k U_{k-1}&=\prod_i \exp\qty(-i\frac{\pi}{2}U_{k-1}^\dagger \qty(\sum_\mu \alpha_{\mu,k}S_i^\mu) U_{k-1})\nonumber\\&=\prod_i \exp\qty(-i\frac{\pi}{2}\sum_\mu \beta_{\mu,k}S_i^\mu),
\end{align}
This implies that $\vec{\alpha}_k=\sum_\mu \alpha_{\mu,k}\hat{e}_{\mu}$, the rotation axis in the lab frame, transforms to $\vec{\beta}_k$ in the toggling frame:
\begin{align}
\sum_\mu U_{k-1}^\dagger (\alpha_{\mu,k} S_i^{\mu}) U_{k-1}=\sum_\mu \beta_{\mu,k}S_i^{\mu}.
\label{eq:rotbeta}
\end{align}
An alternative, more formal way to see this is that $\beta_{\mu,k}$ satisfies the expression
\begin{align}
\exp\qty(i\frac{\pi}{2}\sum_\mu\beta_{\mu,k}S_i^\mu)\qty(\sum_\nu F_{\nu,k}S_i^\nu)\exp\qty(-i\frac{\pi}{2}\sum_\mu\beta_{\mu,k}S_i^\mu)=\sum_\mu F_{\mu,k+1}S_i^\mu.
\end{align}
Using the Baker-Campbell-Hausdorff formula, the above expression can be simplified  to
\begin{align}
i\qty[\sum_\nu\beta_{\nu,k}S_i^\nu,\sum_\mu F_{\mu,k}S_i^\mu]=\sum_\mu F_{\mu,k+1}S_i^\mu,
\end{align}
which gives the same result as Eq.~(\ref{eq:defbeta}).
Summing the total contributions of all imperfect rotations, and using Eq.~(\ref{eq:rotbeta}), we obtain the error Hamiltonian, $\delta H_\text{avg}^\text{rot}$, due to the rotation angle errors
\begin{align}
\delta H_\text{avg}^\text{rot}&=\frac{1}{T}\sum_{k=1}^n t_p \delta \tilde{H}_{P_k} \nonumber\\
&=\frac{1}{T} \sum_{i,\mu} \epsilon_i \qty(\sum_{k=1}^n \beta_{\mu,k}S_i^\mu),
\end{align}
which corresponds to Eq.~(25) of the main text.

We can also formulate a more intuitive geometric picture of the preceding derivation, by considering the Bloch sphere picture of interaction picture spin operators. To better understand this, let us return to the Bloch sphere representation shown in Fig.~2(a-c), lower panels of the main text. We shall call the coordinate system generated by the initial orientation of the $\hat{x}$, $\hat{y}$, $\hat{z}$ axes the fixed \textit{external} coordinate system, and the coordinate system generated by the transformed axis directions as the \textit{toggling-frame} coordinate system. Physically, the axis direction of the toggling-frame coordinate system that points along the $\hat{\mu}$-axis of the external coordinate system is the transformed $\tilde{S}^\mu$ operator: for example, in Fig.~2(b), the second Bloch sphere has the $\hat{y}$ axis pointing up along the $\hat{z}$ direction, indicating that $\tilde{S}^z(t)=U_{k-1}^\dagger S^zU_{k-1}=S^y$ at this time point. 

With this picture in mind, we can now pictorially understand the above technical derivation. At the beginning of the sequence, we start with a toggling-frame coordinate system that coincides with the external reference coordinate system, as shown in the first Bloch sphere of Fig.~2. We then apply rotations along the axes of the fixed \textit{external} coordinate system, which transform the spin operators and rotate the toggling-frame coordinate system.  To specify the axis of rotation in the toggling frame however, we should conjugate by all of the preceding pulses; this corresponds to following the rotations into the toggling-frame coordinate system, and asking which axis the rotation is being applied around in this coordinate system. Since we know the initial and final frame directions in the toggling-frame coordinate system, the rotation effect is uniquely specified and can be easily characterized by the chirality of the rotation connecting the initial and final frames.

With these considerations, we have explicitly shown that cancellation of rotation angle errors is well-captured by the chirality of each frame change, given by $\beta_{\mu,k}$. Note that this applies to both a uniform rotation angle error as well as cases with a field inhomogeneity.

\subsection{Analysis of Rotation Axis Errors}
\label{suppsec:axis}
While we have focused on a particular set of finite pulse effects and pulse imperfections that we believe to be dominant in most experimental systems, our analysis can be readily generalized to other types of imperfections~\cite{rhim1974analysis}. As an example, here we illustrate how to incorporate into our framework the effects of a rotation axis error, in which the applied rotation axis of e.g. an $\hat{x}$-rotation deviates from the designated direction. 

To model this effect, we treat axis errors as a modification of the rotation $\exp[-i\pi S^x/2]$ into $\exp[-i\pi (S^x\pm\zeta S^y)/2]$, and correspondingly $\exp[-i\pi S^y/2]$ into $\exp[-i\pi (S^y\mp\zeta S^x)/2]$. We assume that the rotation axis error results in shifts in the same direction for both $\hat{x}$ and $\hat{y}$ axes. This is justified when microwave control pulses are directly synthesized using an arbitrary waveform generator (AWG): Here, while the $\hat{x}$ and $\hat{y}$ axes are 90 degrees phase shifted with respect to each other for pulses of the same length, different pulse transients for pulses of different length (e.g. $\pi/2$ and $\pi$ pulses) may lead to different relative phases between the rotation axes of each pulse type.
Other types of rotation axis errors can also arise in experiments due to e.g. phase distortions of IQ modulators or nonlinear transient behaviors of microwave amplifiers. We leave the treatment of these more general errors to future work.

We rewrite the rotation axis deviation as the cross product between the $\hat{z}$-axis, corresponding to the disorder direction, and the ideal rotation axis direction. For example, for the rotation $\exp[-i\pi S^x/2]$, we obtain the imperfection contribution $\hat{z}\times \hat{x}=+\hat{y}$. Directly working in the spin operator language, we can make use of the identity
\begin{align}
[S^\mu,S^\nu]=i\epsilon_{\mu\nu\sigma}S^\sigma,
\end{align}
which resembles the form of the cross product. Thus, the rotation axis error Hamiltonian for the pulse $P_k$ can be written as
\begin{align}
\delta H_{P_k} =\sum_i  \frac{\pi\zeta}{2it_p}\qty[S_i^z,\sum_{\nu}\alpha_{\nu,k}S_i^\nu],
\end{align}
where $\vec{\alpha}_k = \sum_\nu \alpha_{\nu,k} \vec{e}_\nu $ is the rotation axis in the lab frame. In the toggling frame, this term becomes
\begin{align}
\delta \tilde{H}_{P_k} &= U_{k-1}^\dagger \delta H_{P_k} U_{k-1} \nonumber \\
&=\sum_i  \frac{\pi\zeta}{2it_p}\qty[\sum_\mu F_{\mu,k} S_i^\mu,\sum_{\nu}\beta_{\nu,k}S_i^\nu],
\end{align}
where we used Eq.~(\ref{eq:Fk-1}) and Eq.~(\ref{eq:rotbeta}).

Utilizing this expression, the average Hamiltonian of the rotation axis error during the $k$-th pulse is calculated as
\begin{align}
\overline{\delta H}_{P_k}
&=\sum_i \frac{\zeta}{it_p}\int_0^{\pi/2}d\theta\qty[\sum_\mu F_\mu(\theta)S_i^\mu,\sum_\nu \beta_{\nu,k}S_i^\nu],
\end{align}
where $F_\mu(\theta)=\cos\theta F_{\mu,k}+\sin\theta F_{\mu,k+1}$ describes the continuous rotation of the frame direction during the pulse. To further simplify the expression, we make use of the fact that $\beta_k$ specifies a rotation from the frame $F_{\mu,k}$ into $F_{\mu,k+1}$, and can thus be written as
\begin{align}
i\sum_\nu \beta_{\nu,k}S^\nu=-\qty[\sum_\lambda F_{\lambda,k}S^\lambda,\sum_\sigma F_{\sigma,k+1}S^\sigma].
\end{align}
This allows us to further simplify the expression as
\begin{widetext}
\begin{align}
\overline{\delta H}_{P_k}&=\sum_i \frac{\zeta}{t_p}\int_0^{\pi/2}d\theta\qty[\sum_\mu (\cos\theta F_{\mu,k}+\sin\theta F_{\mu,k+1})S_i^\mu,\qty[\sum_\lambda F_{\lambda,k}S_i^\lambda,\sum_\sigma F_{\sigma,k+1}S_i^\sigma]]\nonumber\\
&=\frac{\zeta}{t_p}\sum_{i\mu\lambda\sigma}\qty(F_{\mu,k}+F_{\mu,k+1})F_{\lambda,k}F_{\sigma,k+1}\qty[S_i^\mu,\qty[S_i^\lambda,S_i^\sigma]]\nonumber\\
&=\frac{\zeta}{t_p}\sum_{i\mu\lambda\sigma\beta}\qty(F_{\mu,k}+F_{\mu,k+1})F_{\lambda,k}F_{\sigma,k+1} (\delta_{\beta\sigma}\delta_{\mu\lambda}-\delta_{\beta\lambda}\delta_{\mu\sigma})S_i^\beta\nonumber\\
&=\frac{\zeta}{t_p}\sum_{i\mu\beta}\qty[(F_{\mu,k}+F_{\mu,k+1})F_{\mu,k}F_{\beta,k+1}S_i^\beta-(F_{\mu,k}+F_{\mu,k+1})F_{\beta,k}F_{\mu,k+1}S_i^\beta]\nonumber\\\
&=\frac{\zeta}{t_p}\sum_{i\beta} \qty(F_{\beta,k+1}-F_{\beta,k})S_i^\beta,
\end{align}
\end{widetext}
where in the third line, we used the identity
\begin{align}
&\qty[S^\mu,\qty[S^\lambda,S^\sigma]]=\qty[S^\mu,\sum_\nu i\epsilon_{\lambda\sigma\nu}S^\nu]=\sum_{\beta\nu}i^2\epsilon_{\nu\beta\mu}\epsilon_{\nu\lambda\sigma}S^\beta =-\sum_\beta(\delta_{\beta\lambda}\delta_{\mu\sigma}-\delta_{\beta\sigma}\delta_{\mu\lambda})S^\beta,
\end{align}
in the fourth line, we performed the summation over $\lambda$ and $\sigma$ to eliminate the Kronecker deltas, and in the fifth line, we used the fact that at times $k$ and $k+1$, the frame orientation is different, and thus $F_{\mu,k}\cdot F_{\mu,k+1}=0$ for all $\mu$, as well as the fact that $\sum_{\mu}F_{\mu,k}^2=1$. Thus, the condition for rotation axis deviations to be cancelled is that considering all $\pi/2$ pulses (i.e. neglecting $\pi$ pulses), for each axis direction $\hat{x}$, $\hat{y}$, $\hat{z}$, the sum of parities of the frames before each pulse is equal to the sum of parities of the frames after each pulse. Note that if we had included $\pi$ pulses as well, then the sum would fully cancel and become 0, which is consistent with the fact that a global phase rotation of all pulses does not affect the decoupling performance, and can be counteracted by a change of axis.

Incorporating a rotation axis deviation between the $\hat{x}$ and $\hat{y}$ rotations requires modifications to the formalism. This is because the average Hamiltonian rotation axis error term will now depend on the specific pulse physically applied ($\hat{x}$ or $\hat{y}$), which cannot be directly read out in our formalism using only the information of the free evolution frames immediately preceding and following the pulse. One possibility would be to keep track of the full configuration of frame orientations or make use of the history of rotations, as described in Sec.~\ref{suppsec:identity}, though the analysis would be more involved. However, we emphasize that in the case where both $\hat{x}$ and $\hat{y}$ rotations are directly synthesized using an AWG, we expect this type of rotation axis deviation to be very small.

\subsection{Obtaining $S^x$ Transformation Trajectories}
\label{suppsec:identity}
In the main text, we have assumed that the periodic pulse sequence gives rise to an identity evolution operator over one Floquet cycle. However, this need not be the case, since our framework only specifies the transformations of the $S^z$ operator to be periodic, but the transverse $S^{x,y}$ operators may not return to their initial configurations. This may be particularly problematic for sensing purposes, since the phase accumulation in the $\hat{x}$ or $\hat{y}$ directions would then cancel over multiple cycles due to the changing frame directions of $\hat{x}$ and $\hat{y}$ axes. In such situations, one should repeat the frame matrix multiple times until the total unitary rotation by the pulses is identity and use the enlarged frame matrix to define the Floquet period.

Here, we provide a simple method to obtain the evolution of the $S^x$ operator from the frame matrix $\mathbf{F}$ specifying the transformations of $S^z$, such that one can easily keep track of the total unitary rotation up to any point. This is not only useful to determine whether the total unitary rotation is identity, but also helps to directly write down the rotation pulses that need to be applied in the lab frame to realize the transformations specified by $\mathbf{F}$. 

To intuitively understand how the criteria are obtained, let us consider a generic frame orientation in the free evolution period $k$, with the spin operators transformed into $S^z \rightarrow \tilde{S}^z_k$ and $S^x \rightarrow \tilde{S}_k^x$ respectively. We now consider the Bloch sphere intuition developed in Sec.~\ref{suppsec:angleerror}. If $\tilde{S}_{k+1}^z=\pm \tilde{S}_k^x$, this implies that the rotation $P_k$ should be applied along the $\hat{y}$ axis to bring the toggling frame operator pointing along the $\hat{x}$ direction into the $\hat{z}$ direction. Correspondingly, it will also rotate the $\hat{z}$ direction into the $\hat{x}$ direction, with an additional negative sign. Thus, whenever $\tilde{S}_{k+1}^z=\pm \tilde{S}_k^x$, the $\hat{x}$ and $\hat{z}$ operators will switch in the subsequent frame, with the product of the signs of $\tilde{S}_k^x$, $\tilde{S}_{k+1}^x$, $\tilde{S}_k^z$, $\tilde{S}_{k+1}^z$ being $-1$. On the other hand, if $\tilde{S}_{k+1}^z\neq\pm \tilde{S}_k^x$, then the rotation $P_k$ should be applied along the $\hat{x}$ axis, and $\tilde{S}_{k+1}^x=\tilde{S}_k^x$.

With this information, we can sequentially construct the matrix $\mathbf{G}$ that corresponds to a similar representation as $\mathbf{F}$, but for the transformed operator $\tilde{S}^x$. The rules are:
\begin{enumerate}
\item The first column is $\vec{G}_1 = \begin{pmatrix}1\\0\\0\end{pmatrix}$.
\item For each subsequent column $k+1$,  compare the location of the nonzero element in $\vec{F}_{k+1}$ to $\vec{G}_k$. If they are different, copy $\vec{G}_k$ into $\vec{G}_{k+1}$; otherwise, copy $\vec{F}_k$ into $\vec{G}_{k+1}$, and multiply $-1$ if the nonzero elements of $\vec{F}_{k+1}$ and $\vec{G}_k$ have the same sign.
\end{enumerate}
For example, looking at the WAHUHA sequence (see Fig.~2(e) of the main text) we have
\begin{align}
\begin{pmatrix}
{\bf F}\\
\boldsymbol{\tau}
\end{pmatrix}_\text{WAHUHA}
=\begin{pmatrix}
0 & 0 & +1 & 0 & 0\\
0 & +1 & 0 & +1 & 0\\
+1 & 0 & 0 & 0   & +1\\
\tau& \tau& 2\tau& \tau& \tau
\end{pmatrix},\nonumber
\end{align}
from which we can construct
\begin{align}
\mathbf{G}_\text{WAHUHA}
=
\begin{pmatrix}
+1 & +1 & 0 & +1 & +1\\
0 & 0 & -1 & 0 & 0\\
0 & 0 & 0 & 0 & 0\\
\end{pmatrix},
\end{align}
specifying the transformations of the $\tilde{S}^x$ operator. With the known $\tilde{S}^x$ and $\tilde{S}^z$ operators, the $\tilde{S}^y$ operator is uniquely determined by Eq.~(B4) of the main text.

\subsection{Numerical Calculation of Higher-Order Contributions}
\begin{figure}[!ht]
\begin{center}
\includegraphics[width=0.5\columnwidth]{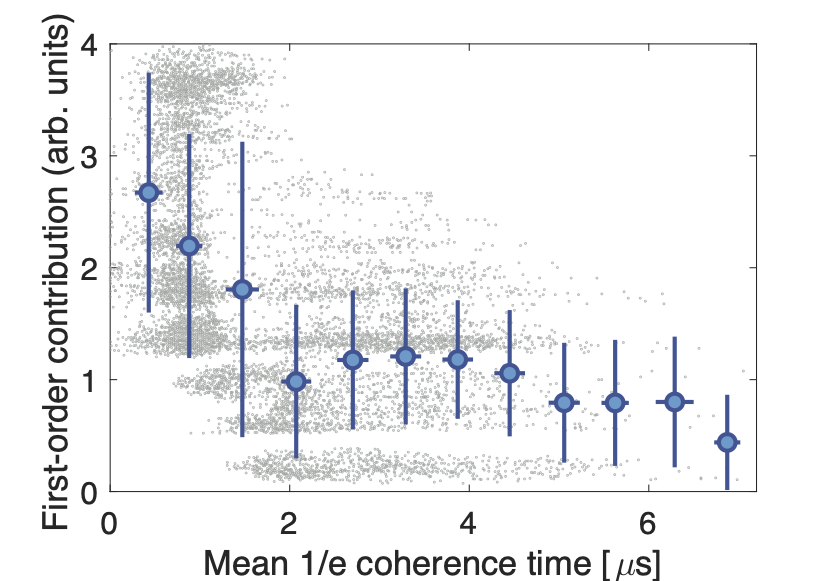}
\caption{Magnitude of first-order Magnus contribution as a function of coherence time for realistic experimental parameters. The scattered points in the background correspond to different Floquet sequences exhibiting different decoherence characteristics. The blue data markers with errorbars denote the mean and standard deviations of binned scatter points. Details of the numerical simulations are given in main text.}
\label{fig:fig1}
\end{center}
\end{figure}

In order to understand the next leading contribution to the effective Hamiltonian beyond the zeroth-order Magnus expansion, we analytically calculate the first-order terms in Eq.~(27,28) of the main text for all 14,080 sequences introduced in Sec.~IV.2., including all finite pulse duration effects at this order. We then estimate the numerical strength of each contribution based on realistic experimental system parameters introduced in the main text, and plot the variance of the magnitude of the first-order Magnus expansion in Fig.~\ref{fig:fig1}. As one can see, there is a strong correlation between the magnitude of the first-order term and the coherence, confirming that higher-order contributions cause corrections to the coherence decay, and they can be systematically understood using our formalism.

\subsection{Engineering Hamiltonians for Quantum Simulation}
\label{suppsec:simulation}
In this section, we provide additional details of engineering Hamiltonians with different disorder strengths and interaction forms for the purpose of quantum simulation. The main idea is that instead of choosing toggling-frame evolution times to fully cancel the contributions in Rules 1-4 in Tab.~1 of the main text, we engineer them to have a finite value, thus resulting in a desired target Hamiltonian. We shall restrict our attention to a subset of the achievable Hamiltonians, which show very rich behavior in terms of the range of accessible thermalization properties.

By introducing imbalanced time evolutions in the toggling frames, we can generate a tunable vector disorder field $\vec{h}_\text{eff}$ for driven qubits, described by 
\begin{align}
	(\vec{h}_\text{eff})_i =  \sum_{\mu=x,y,z} \hat{e}_\mu \left[  \frac{h_i}{T}  \sum_{k=1}^n F_{\mu,k}\qty (\tau_k + \frac{4}{\pi}t_p) \right],
 \end{align}
where $h_i$ is the original on-site disorder strength at site $i$, with the corresponding engineered disorder Hamiltonian
\begin{align}
	H_\text{avg}^\text{dis} = \sum_i (\vec{h}_\text{eff})_i \cdot \vec{S}_i.
\label{eq:Hdisnew}
\end{align}The disorder can thus have both longitudinal and transverse components in the toggling frame. Moreover, many-body interactions can also be tailored in a similar fashion. As an example, for systems with Ising and symmetric spin-exchange interactions, their original interaction Hamiltonian, $H_\text{int} = \sum_{ij} [J_{ij}^S (S_i^x S_j^x + S_i^y S_j^y) + J_{ij}^I S_i^z S_j^z]$, can be transformed to
\begin{align}
	H_\text{avg}^\text{int} = \sum_{ij} [\tilde{J}_{ij}^S (S_i^x S_j^x + S_i^y S_j^y) + \tilde{J}_{ij}^I S_i^z S_j^z]
\label{eq:Hintnew}
\end{align}
with tunable interaction strengths
\begin{align}
	\tilde{J}_{ij}^S & = \frac{1+c}{2}J_{ij}^S + \frac{1-c}{2}J_{ij}^I, \\
	\tilde{J}_{ij}^I &= (1-c) J_{ij}^S + c J_{ij}^I. 
\end{align}
Here, the interaction-type tuning coefficient $c$ is given by
\begin{align}
	c = \frac{1}{T} \left[ \sum_{k=1}^n |F_{z,k}| (\tau_k + t_p) \right]
\end{align}
with the constraint $\sum_{k=1}^n |F_{x,k}| (\tau_k + t_p) = \sum_{k=1}^n |F_{y,k}| (\tau_k + t_p)$. Note that the interaction tuning coefficient $c$ will also impact the achievable effective disorder strength, since the effective disorder along the $\hat{\mu}$ axis is limited by the proportion of evolution time along this axis. Thus, the maximal achievable disorder for given $c$ will be $\max{(\vec{h}_\text{eff})} = [\frac{1-c}{2},\frac{1-c}{2},c]$, which results in $W_\text{eff} =W \sqrt{c^2+(1-c)^2/2}$ where $W$ and $W_\text{eff}$ correspond to the original and effective disorder strengths, respectively. 

Combining these techniques with the other rules in Tab.~1 for robust Hamiltonian engineering, we can engineer a wide range of Hamiltonians. Importantly, the disorder and interaction Hamiltonians [Eqs.~(\ref{eq:Hdisnew},\ref{eq:Hintnew})] can be independently engineered due to their different functional dependence on $F_{\mu,k}$ (linear for $H_\text{avg}^\text{dis}$ and quadratic for $H_\text{avg}^\text{int}$). In addition, one can also engineer uniform single-body terms via intentional detuning or systematic rotation angle deviations (rule 4), and disordered XY-type interactions via incomplete cancellation of cross terms (rule 3). By utilizing higher-order Magnus expansions, one can also engineer effective three-body interactions generated by the commutators. This allows one to obtain the desired many-body Hamiltonian with tunable disorder and interaction type. As discussed in the main text, taking the case of NV centers as an example where $J^S_{ij}=-J^I_{ij}$ and $J_{ij}^A=0$, we can engineer Ising ($c=0$), Heisenberg ($c=1/3$), XY ($c=1/2$), and dipolar-like ($c=1$) interactions. These can exhibit very different thermalization behavior, as the Ising interaction limit is expected to be integrable, while the exchange interactions can facilitate spreading of excitations.

There are, however, a few constraints on the form of the engineered Hamiltonian: First, the Heisenberg component, $\vec{S}_i \cdot \vec{S}_j$, of the interaction will remain invariant, since it is not transformed under global rotations; Second, some of the Hamiltonian terms transform in the same way (for instance disorder and anti-symmetric spin exchange), and consequently cannot be independently engineered; Finally, the resultant Hamiltonian may experience a rescaling in magnitude, due to the finite projection of the initial Hamiltonian onto the final Hamiltonian. Eventually, contributions from higher-order terms in the Magnus expansion may also become important.

\subsection{Proof of Optimal Sensitivity Under Interaction Decoupling Constraints}
\label{suppsec:proofoptimal}
Our formalism also allows us to easily determine the optimal achievable effective field strengths under the condition that interaction effects are decoupled. As described in the main text, the condition for Ising and spin-exchange interaction-decoupling requires equal evolution times along each of the $\hat{x}$, $\hat{y}$, $\hat{z}$ axes in the toggling frame. In the typical scenario where a spin-echo is performed along each axis, an effective sensing field $\vec{B}_\text{eff}=B_\text{XY-8} \qty[\frac{1}{3},\frac{1}{3},\frac{1}{3}] $ is generated at the dominant resonance, with a magnitude
\begin{align}
|\vec{B}_\text{eff}| = B_\text{XY-8}/\sqrt{3}\approx 0.577\cdot B_\text{XY-8},
\end{align}
a factor of $\sqrt{3}$ smaller than the maximal field strength $B_\text{XY-8}$ achievable for the XY-8 sensing sequence, where interactions are not decoupled. In fact, despite the relative reduction in the sensing field magnitude, our sequence is optimal for sensing with disorder-dominated interacting spin systems, as interaction-decoupling requires equal evolution time along different axes, and a fast spin-echo structure is required to not only effectively detect external AC signals but also rapidly suppress the effects of disorder.

If we broaden the scope to include pulse sequences that do not have an immediate echo structure, then the above pulse sequence can be further improved into a version that is provably optimal in the limit of infinitely short pulses. Here, the key observation is that the amount of {\it phase accumulation} along each axis can be imbalanced, even while spending equal amount of \textit{time} along all axes to decouple interactions. To theoretically derive the maximum AC sensitivity under the constraint of interaction decoupling (where the evolution time along each axis is required to be equal), the sinusoidal amplitude modulation of the target sensing signal needs to be carefully taken into account. The effective field strengths along each axis at the resonance are proportional to the weighted integrals of the sinusoidal target signal, and the sum of amplitudes is constrained by the identity
\begin{align}
\qty|B_x|+\qty|B_y|+\qty|B_z|\leq B_\text{XY-8},
\label{eq:Bineq}
\end{align}
where $B_\text{XY-8}$ is the effective field strength obtained from the conventional XY-8 sequence. At the same time, the best AC sensitivity is achieved when the total effective field strength $\qty|\vec{B}_\text{eff}|=\sqrt{B_x^2+B_y^2+B_z^2}$ is maximized. To optimize $|\vec{B}_\text{eff}|$, we first assume that the inequality Eq.~(\ref{eq:Bineq}) is saturated, in which case $\qty|B_x|=B_\text{XY-8}-\qty|B_y|-\qty|B_z|$. Without loss of generality, let us also assume $B_x>B_y>B_z>0$. Taking the partial derivative, we obtain
\begin{align}
\frac{\partial |\vec{B}_\text{eff}|^2}{\partial B_z}
&=4B_z-2B_y-2B_\text{XY-8}<0.
\end{align}
Since $B_z<B_y<B_\text{XY-8}$, we see that it is desirable to minimize the smallest element. We thus hold $B_z$ fixed at its minimal value, and optimize the values of $B_x$ and $B_y$, in which case
\begin{align}
\frac{\partial |\vec{B}_\text{eff}|^2}{\partial B_y}=2B_y-2B_x<0.
\end{align}
Similarly, the effective field strength is maximized when $B_y$ is minimized under the constraint that $B_z$ is minimal. Given this understanding, optimal sensitivity will be achieved when the phase accumulation under the sinusoidal target signal is distributed along the three axes to maximize their difference. This implies that it is desirable to distribute all phase accumulation along the $\hat{x}$-axis to be near the anti-nodes of the sinusoid and phase accumulation along the $\hat{z}$-axis to be near the nodes, as illustrated in Seq.~I in Fig.~8 of the main text. Using Eqs.~(36,37) presented in the main text, we estimate the effective sensing field of Seq.~I to be $\vec{B}_\text{eff} = B_\text{XY-8} \qty[1-\frac{\sqrt{3}}{2},\frac{\sqrt{3}}{2}-\frac{1}{2},\frac{1}{2}]$ in the perfect pulse limit with zero pulse duration. The corresponding sensing field magnitude $|\vec{B}_\text{eff}|\approx 0.634B_\text{XY-8}$ is the theoretical upper limit of the effective field strength under the constraint of interaction decoupling, about 10$\%$ improved over fast-echo-based sensing sequences, such as Seq.~B, that require an even distribution of phase accumulation between the three axes. For realistic cases of finite pulse duration, we can also use the same strategy to optimize the effective field strength, by maximizing the phase accumulation imbalance between different axes. In such cases, we can choose to reduce the number of short intermediate frames along the $\hat{z}$ axis, and instead satisfy the interaction decoupling requirements by increasing the duration of free evolution time along the $\hat{z}$ axis. This will further increase the difference between the phase accumulation along different axes, leading to high sensitivity for the case of finite pulse duration. 

\section{Experimental Details}
\label{suppsec:experiment}

\subsection{Experimental Setup}
Our experimental system consists of a dense electronic spin ensemble ($\sim$15 ppm) of nitrogen-vacancy (NV) centers in diamond. We utilize a static external magnetic field along the [1,1,0]-direction to split the spin-1 ground states $\ket{0,\pm1}$ and resonantly address the $\ket{0}$ and $\ket{-1}$ levels, effectively isolating an ensemble of two-level systems consisting of NVs with two groups of crystallographic axis orientations. These subgroups of NVs ($\sim$8 ppm) are characterized to have large on-site disorder ($W\sim(2\pi)~4.0$ MHz) originating from paramagnetic impurities and inhomogeneous strain, with modest interaction strengths through long-range magnetic dipolar interactions ($J\sim(2\pi)~35$ kHz within a single group of NVs with the same axis orientation). Note that the form of the interaction between two resonant NV groups involves Ising, symmetric and anti-symmetric spin exchange interactions with complex, position-dependent coefficients~\cite{kucsko2018critical}. The {\it intrinsic} magnetic dipolar interaction strength is significantly larger than {\it extrinsic} decoherence rates due to coupling to the environment. Indeed, we have verified that the coherence time of the standard XY-8 sequence is limited by NV-NV interactions~\cite{choi2017depolarization,kucsko2018critical,choi2019probing}. We utilize optical excitation and readout to initialize and detect the spin states and use resonant microwave driving to manipulate the spin state. The microwave pulses are directly synthesized using a high sampling rate AWG (Tektronix 7122B), amplified through a microwave amplifier (Mini Circuits ZHL-16W-43-S+), and delivered to the NV ensemble with an $\Omega$-shaped coplanar waveguide. We cut out a nanobeam from a bulk sample using an angled etching procedure~\cite{Burek2012free}, to improve optical and microwave control homogeneity. 

\subsection{Data Analysis}
We now discuss the detailed forms of fitting performed for the experimental data. For the spin coherence decay measurements in Fig.~8(b), we fix the free evolution time $\tau=15$ ns and $\pi/2$ pulse length $t_p = 6$ ns, and sweep the number of times the sequence is repeated. We fit the experimental data to a stretched exponential profile $y(t)=y_0 \exp[-(t/T_2)^\alpha]$, with fitting parameters being the $1/e$ coherence time $T_2$, the exponent $\alpha$ and the maximum contrast $y_0$. For ease of comparison, we plot the normalized decay curves defined as $y(t)/y_0$.

In the rotation error robustness measurements in Fig.~8(c) of the main text, we fit the experimental data to a stretched exponential with a finite contrast sinusoidal modulation: 
\begin{align}
y(t)=y_0 \exp[-(t/T_2)^\alpha]\cdot (A \cos(2\pi f t)+1-A),
\end{align}
where $A$ and $f$ characterize the contrast modulation depth and frequency, respectively. In Fig.~8(d), we extract $f$ as a function of the Rabi frequency $\Omega$, and we introduce the systematic rotation error $\epsilon = (\Omega - \Omega_0) t_p$, where $\Omega_0 = \pi/2t_p$ is the precalibrated Rabi frequency. Taking the mean of the extracted $f$ values averaged over different initial states, we fit the profile of $f$ to the functional form $f(r) \propto |r-r_0|$, where $r =(\Omega-\Omega_0)/\Omega_0$ and $r_0$ is a small lateral offset.

\bibliography{SensingTheory_supp_bibtex}